\documentclass[mnsc,nonblindrev]{informs3_hide}
\OneAndAHalfSpacedXI 

\usepackage[english]{babel}
\usepackage[autostyle, english = american]{csquotes}
\MakeOuterQuote{"}
\usepackage[colorlinks,linkcolor=blue, citecolor=blue]{hyperref}
\usepackage[normalem]{ulem}

\usepackage{cleveref}
\crefname{subsection}{section}{subsections}
\usepackage{eqnarray}
\usepackage{algorithm,algpseudocode}
\usepackage{amsmath, bbm, enumitem, xparse, bm, lipsum}
\usepackage{amssymb}

\newcommand{\R}{\mathbb{R}}
\newcommand{\bB}{\bm{B}}
\newcommand{\bA}{\bm{A}}
\newcommand{\bK}{\bm{K}}
\newcommand{\bL}{\bm{L}}
\newcommand{\bI}{\bm{I}}
\newcommand{\ba}{\bm{a}}
\newcommand{\bp}{\bm{p}}
\newcommand{\bM}{\bm{M}}
\newcommand{\bH}{\bm{H}}
\newcommand{\bD}{\bm{D}}
\newcommand{\bv}{\bm{v}}

\newcommand{\bx}{\bm{x}}

\newcommand{\bz}{\bm{z}}
\newcommand{\bw}{\bm{w}}

\newcommand{\bG}{\bm{G}}

\newcommand{\bQ}{\bm{Q}}

\newcommand{\bY}{\bm{Y}}

\usepackage[dvipsnames]{xcolor}

\usepackage{xparse}

\NewDocumentEnvironment{myproof}{o}
{\IfNoValueTF{#1}{\paragraph{{Proof.} }} {\paragraph{{#1.} }} }
{\hfill$\Halmos$}

\usepackage{natbib,bbm,multirow,multicol}
 \bibpunct[, ]{(}{)}{,}{a}{}{,}%
 \def\bibfont{\small}%
\TheoremsNumberedThrough     
\ECRepeatTheorems

\EquationsNumberedThrough    

\begin{document}


\RUNAUTHOR{}

\RUNTITLE{}

\TITLE{\Large Decentralized Multi-product Pricing:
Diagonal Dominance, Nash Equilibrium, and Price of Anarchy
}

\ARTICLEAUTHORS{%
\AUTHOR{$\text{Boxiao Chen}^{\dagger}$, $\text{Jiashuo Jiang}^{\ddagger}$, $\text{Stefanus Jasin}^{\S}$}

\AFF{\ \\
$\dagger~$Naveen Jindal School of Management, The University of Texas at Dallas\\
$\ddagger~$Department of Industrial Engineering \& Decision Analytics, The Hong Kong University of Science and Technology\\
$\S~$Department of Technology and Operations, Ross School of Business, University of Michigan
}
}

\ABSTRACT{Decentralized pricing in multi-product firms and marketplaces
can lose revenue because each decision maker ignores how its price
affects the demand for other products. We quantify this loss through
the price of anarchy of a game in which the price of each product is
set to maximize that product's own revenue. Demand is linear, and
cross-price effects may be substitutes or complements and need not be
symmetric. Under a two-sided diagonal dominance condition with
parameter $\mu\in[0,1)$, which bounds the aggregate absolute cross-price
effects in each row and column relative to the own-price effect, we
show that the game has a unique Nash equilibrium and that equilibrium
revenue is at least a fraction $4(1-\mu)/(2-\mu)^2$ of the centralized
optimum. This bound was previously known for symmetric gross
substitutes; we show that it remains valid for signed and asymmetric
interactions, and that it is tight. Because asymmetry breaks the
simultaneous diagonalization underlying existing analyses, we instead
evaluate the revenue gap at the equilibrium itself. This yields an
exact formula for the worst-case price of anarchy of each demand matrix
in terms of a normalized interaction matrix, which can be much sharper
than the scalar bound in heterogeneous interaction networks. The same
argument extends the guarantee to jointly concave nonlinear demand
under pointwise dominance of the demand Jacobian, for every
pure-strategy equilibrium.}

\KEYWORDS{Price of Anarchy, Decentralized Pricing, Nash Equilibrium, {Asymmetric Cross-Price Effects}}


\maketitle

\section{Introduction}\label{sec:Intro}

Pricing strategies form the backbone of modern revenue
management and play a central role in profitability across a wide range
of industries, including retail, e-commerce, logistics, and digital
services.
For firms that manage large portfolios of interconnected products, the
conceptual benchmark for decision making is a
\emph{centralized pricing scheme}.
Under this paradigm, a firm with perfect information and unlimited
computational capability jointly optimizes the prices of all $N$
products, fully accounting for the network of substitution and
complementarity effects that link demand across the portfolio.
By internalizing these cross-product externalities—such as the fact
that discounting a flagship product may cannibalize sales of a premium
alternative or stimulate demand for a complementary accessory—the
centralized approach is guaranteed, in principle, to maximize total
revenue.

Despite its conceptual appeal, centralized pricing is
often unattainable in practice, and pricing decisions are commonly
decentralized, either by design or by necessity, with individual
products or teams setting prices based on local objectives and
information. Such decentralized decision making generally fails to
internalize cross-product effects and can therefore lead to outcomes
that diverge from the centralized optimum.
In this paper, we investigate the performance of a
\emph{decentralized pricing scheme} relative to the centralized
benchmark.
We model decentralized pricing as a noncooperative game in which the
price of each product is chosen to maximize its own revenue, taking the
prices of all other products as given.
Our analysis focuses on the resulting Nash equilibrium and on
quantifying the efficiency loss relative to the centralized optimum—a
metric commonly referred to in the literature as the
\emph{Price of Anarchy} (PoA).

Our emphasis on decentralized pricing and its associated Nash
equilibrium is motivated by two observations: the
centralized optimum is often unattainable in practice, and the Nash
equilibrium is a natural outcome in many applications, particularly
when full demand information is unavailable. We illustrate these
observations in two settings.

\textbf{1. Algorithmic complexity and learning.}
Even when a single firm controls all $N$ products, fully centralized
pricing requires accurate knowledge of the entire demand system: the
$N\times N$ matrix of cross-price effects or, for nonparametric models,
$N$ demand functions of $N$ variables each, whose estimation is subject
to the curse of dimensionality. Estimating these interactions
can be statistically challenging, particularly in high-dimensional
settings with limited data, even when the resulting centralized
optimization problem is tractable. Learning the centralized optimum
also requires pooling data across all products and coordinating price
experiments in all $N$ dimensions. As a result, firms commonly deploy
simplified or modular learning algorithms, such as product-level
learning modules, that treat products as independent. Each such module
uses only its own product's bandit feedback, such as realized revenue,
and adjusts the product's price based on the estimated response of that
revenue to its own price, thereby bypassing the need to estimate any
cross-price effect or the product's demand as a function of all $N$
prices. The feedback each module receives, however, depends on the
prices chosen by all other modules. From the perspective of a single
product, the other prices act as part of a changing environment, and
each module keeps adjusting its price as long as doing so increases its
own revenue. Any stable outcome of this process is therefore a price
vector from which no product can gain by unilaterally changing its
price, that is, a Nash equilibrium. Indeed, under the diagonal
dominance condition imposed in this paper, the resulting pricing game
is strongly monotone, and independent gradient-based and no-regret
algorithms driven by each product's own feedback provably converge to
its unique Nash equilibrium \citep{ba2025doubly, jordan2025adaptive}.
The firm thus reaches the Nash equilibrium without any module, or the
firm as a whole, ever knowing the full demand system.

\textbf{2. Multi-agent competition.}
Many contemporary marketplaces are inherently fragmented. On online
retail platforms and in decentralized supply chains, products are
frequently sold by independent agents, third-party sellers, or distinct
business units within a larger organization. These agents maximize the
revenue of their own products rather than the aggregate revenue of the
platform or firm, so the Nash equilibrium is the natural outcome of
their interaction. When demand is unknown, each agent observes only its
own revenue data and has neither access to other agents' data nor
control over their prices, so estimating the full demand system is not
merely costly but infeasible. Reaching the Nash equilibrium requires no
such information: under the same learning dynamics as above, each agent
needs only to learn how its own revenue responds to its own price at
the current prices, and the joint prices still converge to the Nash
equilibrium \citep{ba2025doubly, jordan2025adaptive}. The price of
anarchy then bounds the revenue that the platform could gain by
coordinating sellers' prices, for example through pricing tools or
price recommendations, and hence, under revenue-based commissions, the
platform's own fee income.

\vspace{1mm}
Given these practical considerations, it is essential to understand the
magnitude of the revenue loss induced by decentralized pricing.
The price of Anarchy has been studied extensively in other
settings—most notably in transportation networks (e.g.,
\cite{youn2008price}) and routing games (e.g.,
\cite{colini2020selfish}).
In differentiated-product price competition, \citet{farahat2009profit}
establish sharp profit-efficiency bounds under affine, symmetric, gross-substitute
demand. However, in practice, the cross-price interactions are typically signed (substitute or complementary) and asymmetric.
This perspective motivates the central question of our study:

\vspace{2mm}
\begin{center}
    \emph{How much revenue is lost due to decentralized pricing, relative
    to the centralized optimum, when
    cross-price effects exist and may be signed and asymmetric?}
\end{center}

\vspace{2mm}
To address this question, we consider the setting with a linear demand model
under structural assumptions that ensure
well-posedness.
Within this framework, we characterize both the centralized optimal
prices and the decentralized Nash equilibrium in closed form, and we
quantify the resulting efficiency loss using the Price of Anarchy
(PoA), defined as the ratio of revenue achieved under decentralization
to that under centralized optimization.
Our focus is the development of efficiency guarantees to signed and
potentially asymmetric interactions under two-sided absolute
diagonal dominance. We also identify conditions under
which the same revenue comparison continues to hold beyond linear demand.

\subsection{Our Main Results and Contributions}

We consider a linear demand model in which the demand for each
product depends linearly on the prices of all products in the
portfolio.
Within this framework, we rigorously analyze the ratio between the
aggregate revenue generated at the Nash equilibrium of the decentralized
pricing game and the maximum revenue achievable under joint
optimization by a centralized decision maker.
Our main results and their relation to prior work are as follows.

\vspace{1mm}
\textbf{Existence, uniqueness, and closed-form characterization.}
As a basis for the efficiency analysis, we first establish the basic
structural properties of the decentralized pricing game.
Under the stated assumptions, specifically, that the demand
sensitivity matrix satisfies two-sided strict diagonal dominance,
without requiring symmetry, we
prove that the game admits a unique pure-strategy Nash equilibrium.
Beyond existence and uniqueness, we derive explicit closed-form
expressions for the equilibrium price vectors under both centralized
and decentralized pricing.
These characterizations allow us to explicitly isolate the distortion
introduced by decentralization.
In particular, we show that while the centralized planner sets prices
using the symmetric part of the demand matrix and thus fully internalizes all
cross-product effects,  decentralized agents effectively operate on a modified matrix that downweights off–diagonal
interactions, leading to a systematic misalignment of incentives across products.

\vspace{1mm}
\textbf{Worst-case efficiency bounds via diagonal dominance.}
A central challenge in Price of Anarchy analysis is to obtain bounds
that are robust across a wide range of market structures.
We show that the competition-intensity bound of
\citet{farahat2009profit} continues to hold for signed and asymmetric
interactions under our two-sided dominance condition. Because
asymmetric interactions rule out the simultaneous diagonalization
underlying their analysis, our proof instead evaluates the revenue gap
at the equilibrium itself, using only the equilibrium conditions. The bound depends
on a single scalar parameter $\mu \in [0,1)$.
This parameter quantifies the degree of diagonal dominance of the demand
matrix and can be interpreted as the aggregate absolute strength of cross-price
effects, whether substitution or complementarity, in both rows and columns, relative to own-price
sensitivities.
We prove that for any market satisfying this dominance condition, the
ratio of decentralized to centralized revenue is bounded below as follows:
\[
\text{PoA} \;\ge\; \frac{4(1-\mu)}{(2-\mu)^2}.
\]

\vspace{1mm}
\noindent
This bound {shows how a worst-case efficiency guarantee depends on
the intensity of product interdependence.}
When products are highly differentiated and cross-effects are weak
($\mu \approx 0$), the bound approaches one, implying that decentralized
pricing is nearly optimal.
{As $\mu$ approaches one, the worst-case guarantee deteriorates.
The actual revenue loss for a particular matrix also depends on the
signs and direction of its interactions.}

\vspace{1mm}
\textbf{Tightness of the bound.}
We further examine whether the derived bound is merely conservative or
whether it represents a fundamental limit.
We show that the bound {remains tight because of an explicit worst–case market structure.}
In particular, we analyze a fully symmetric, exchangeable model in
which each product interacts with every other product with equal
intensity.
In this setting, we show that the realized {revenue ratio equals}
the {stated} lower bound.
This establishes that no sharper bound depending solely on the diagonal
dominance parameter $\mu$ can be obtained {uniformly over the present
unrestricted affine class.}

\vspace{1mm}
\textbf{Exact spectral characterization of the Price of Anarchy.}
We further observe that the scalar parameter $\mu$ reflects a
worst-case aggregation of cross-product effects and may therefore be
overly conservative in sparse or heterogeneous interaction networks.
To address this limitation, we develop an exact spectral
characterization of the {worst-case Price of Anarchy over all nonzero
real intercept vectors, for each fixed signed and potentially asymmetric
demand matrix in the unrestricted affine model.}
We show that the efficiency loss is determined precisely by the
spectral properties of a symmetric positive definite revenue matrix.
Equivalently, the exact loss is determined by a symmetric positive
semidefinite matrix built from the normalized directed interactions
and the symmetric part of the demand matrix. The largest singular
value of the normalized interaction matrix gives a convenient lower bound.
This characterization yields instance-specific efficiency guarantees
that explicitly account for the topology of the product interaction
network.
As a result, it can substantially improve upon the worst-case bound in
settings such as hub-and-spoke or other heterogeneous market
structures.

\vspace{1mm}
\textbf{Extension to nonlinear demand.}
Finally, we show that the same revenue guarantee holds for {smooth, nonnegative, }jointly
concave nonlinear demand on a common nonnegative price box, under
two-sided diagonal dominance of the price-dependent Jacobian. The
comparison applies to every pure-strategy equilibrium, including
boundary equilibria, without imposing symmetry. A separate
marginal-revenue condition gives equilibrium uniqueness, and a
curvature-adjusted bound permits controlled departures from demand
concavity.

\vspace{1mm}
Taken together, our results {extend established revenue-efficiency
benchmarks to signed and asymmetric multi-product demand systems and
to a class of nonlinear demand functions.}
They bridge the pricing literature with broader work on efficiency loss
in games and decentralized optimization, and they offer concrete
insights into when decentralized pricing is nearly optimal and when
{additional coordination can improve revenue.}

\subsection{Further Related Work}

We now provide an overview of additional related literature.
Broadly speaking, our work is related to three streams of the literature:
(i) the literature on multi-product pricing problems,
(ii) the literature on Price of Anarchy,
and (iii) the literature on Nash equilibrium computation and convergence
analysis.

\vspace{1mm}
\textbf{Multi-product pricing}. 
Multi-product pricing is a fundamental problem in operations research 
and operations management, especially in dynamic and data-driven 
environments. It has been extensively studied within the framework 
of Network Revenue Management (NRM). Since the early works 
(e.g., \cite{gallego1994optimal, talluri1998analysis, 
gallego1997multiproduct}), significant effort has been devoted to 
improving regret guarantees, progressing from $O(\sqrt{T})$ to $O(1)$. 
For example, \cite{reiman2008asymptotically} shows that re-solving 
the ex-ante relaxation once yields an improved regret bound of 
$o(\sqrt{T})$. Subsequently, \cite{jasin2012re} and 
\cite{jasin2014reoptimization} demonstrate that, under a 
non-degeneracy condition, a resolving policy achieves $O(1)$ regret. 
More recently, \cite{bumpensanti2020re} proposes an infrequent 
re-solving policy and establishes an $O(1)$ regret bound without 
the non-degeneracy assumption. Using a different approach, 
\cite{vera2021bayesian} proves the same $O(1)$ upper bound for 
the NRM problem. A variety of extensions have been studied using 
different methodologies (e.g., \cite{jiang2022degeneracy, 
freund2019good, ao2025learning}). Another formulation of multi-product pricing incorporates inventory 
replenishment decisions (e.g., \cite{burnetas2000adaptive, 
chen2021nonparametric, chen2022primal, chen2024optimal}). Finally, 
bandit-based formulations have also been widely adopted to study 
multi-product pricing, leading to a range of approaches under 
different settings (e.g., \cite{besbes2009dynamic, 
keskin2014dynamic, besbes2015surprising, cohen2020feature, 
bu2022context, xu2025joint}). 

{The centralized pricing studies reviewed above assume that a} 
single decision maker sets prices for all products. In contrast, 
our paper studies the performance of a decentralized pricing scheme 
in which each product optimizes its own price.

\vspace{1mm}
\textbf{Price of Anarchy analysis}. 
The concept of the Price of Anarchy (PoA), introduced by 
\citet{koutsoupias1999worst}, quantifies the efficiency loss arising 
from selfish behavior in decentralized systems. Establishing 
worst-case bounds on the PoA under various conditions has been a 
central theme since the seminal work of \cite{papadimitriou2001algorithms}, 
{including efficiency bounds for networks with affine latency functions.} 
These results were subsequently extended in 
a series of works 
\cite{roughgarden2004bounding, correa2004selfish, correa2007fast, 
perakis2007cost, correa2008geometric}. A comprehensive overview can 
be found in the survey by \cite{roughgarden2007routing}.  The PoA has also been extensively studied in the context of supply 
chain competition. Unlike routing games, where agents minimize 
latency, supply chain agents (e.g., retailers and suppliers) 
typically maximize profit, often giving rise to ``tragedy of the 
commons'' phenomena. \cite{perakis2007price} quantified the efficiency 
loss of decentralized supply chains operating under price-only 
contracts. 
\cite{he2017noncooperative} studied the PoA of inventory replenishment 
policies. Another stream of work applies PoA analysis to service 
operations and queueing systems, where customers decide whether to 
join a queue or which server to select based on individual incentives. 
Various results have been obtained across different settings 
(e.g., \cite{haviv2007price, hassin2016rational, gaitonde2023price}); 
see also the early monograph by \cite{hassin2003queue}. PoA has 
further been studied in pricing and auction settings 
(e.g., \cite{johari2009efficiency, johari2011parameterized, 
chen2012design, balseiro2019learning}). 


The closest comparison is \citet{farahat2009profit}, who analyze
affine, symmetric, gross-substitute demand with constant marginal costs
and arbitrary fixed multiproduct ownership.
On the common symmetric substitute subclass, our worst-case bounds and
exchangeable tightness construction coincide with theirs. However, our
analysis focuses on product-level pricing with signed and potentially
asymmetric cross-price effects under two-sided absolute diagonal
dominance. For asymmetric interactions, total revenue depends on each pair of
cross-price effects only through their sum, whereas each product's
equilibrium price responds only to how the other prices affect its own
demand. The matrices governing the two revenues therefore no longer
share an eigenbasis, and the simultaneous-diagonalization argument of
\citet{farahat2009profit} does not apply. We instead evaluate the revenue gap at the equilibrium
itself (Section~\ref{sec:spectral}), which retains the full directed
coefficients. Because this argument does not rely on closed-form
equilibrium prices, it also extends to jointly concave nonlinear demand
(Section~\ref{sec:nonlinear}), whereas the analysis of
\citet{farahat2009profit} is restricted to affine demand.

\vspace{1mm}
\textbf{Equilibrium computation and convergence}. 
Our problem is related to the broad literature on learning in games, 
ranging from query complexity in static environments to convergence 
in repeated interactions. A major line of work studies the sample 
complexity of computing Nash equilibria. In settings with deterministic 
observations, \citet{maiti2023query} proposed an algorithm that 
identifies all Nash equilibria by querying only a limited portion of 
the game matrix; this result was further improved by 
\citet{dallant2024finding} and \citet{dallant2024optimal}. Under noisy 
observations, \citet{zhou2017identify} employed a 
lower-upper-confidence-bound (LUCB) approach 
\citep{kalyanakrishnan2012pac} to identify pure-strategy Nash equilibria 
with finite-sample guarantees, which were later sharpened by 
\citet{maiti2024near}. For mixed equilibria, \citet{maiti2023instance} 
established high-probability bounds, while \citet{jiang2025lp} derived 
instance-dependent bounds in expectation using a novel LP-based 
approach. 

Complementary to these query-complexity results is the extensive 
literature on computing Nash equilibria via regret minimization in 
repeated games. \citet{freund1999adaptive} established the classical 
$O(1/\sqrt{T})$ convergence rate. To overcome this slow rate, subsequent 
works \citep{daskalakis2011near, rakhlin2013optimization, 
syrgkanis2015fast, zhang2022no, cai2022tight} developed techniques to 
accelerate either average-iterate or last-iterate convergence. These 
learning guarantees have since been extended beyond normal-form games 
to richer settings, including convex-concave games 
\citep{abernethy2018faster}, strongly monotone games 
\citep{ba2025doubly, jordan2025adaptive}, extensive-form games 
\citep{farina2019online, lee2021last}, and multi-agent reinforcement 
learning. Finally, learning-in-games has also been studied in pricing 
contexts (e.g., \cite{balseiro2019learning, golrezaei2020no, 
yang2020competitive, li2024lego, bracale2025online}), with a variety of 
approaches offering convergence guarantees. 

In contrast, our paper does not focus on designing learning algorithms 
for games. Instead, we study the {Nash equilibrium outcome of the pricing
game, which can arise under suitable decentralized updates,} and evaluate
its performance relative to the globally optimal solution of the pricing
problem.

\section{Problem Formulation and the Centralized Pricing Scheme}
\label{sec:model}

In this section, we discuss the model, the assumptions, and the centralized pricing scheme.

\subsection{The Model}

We consider a firm that sells $N$ products by choosing prices to maximize the reward. To be specific, we denote by $p_{i}$ the price chosen by the firm, for every product
$i\in\{1,\ldots,N\}$.
Let
$\bm{p}=(p_{1},\ldots,p_{N})$
denote the vector of prices. The realized demand for product $i$ is given as
\[
D_{i}(\bm{p})
=
F_i(\bm{p})+\epsilon_{i},
\]
where $F_i(\bm{p})$ is the expected demand as a function
of the entire price vector and $\epsilon_{i}$ is a
zero-mean random shock.
The noise terms $\epsilon_{i}$ are assumed to be
independent between products. {We first assume that expected demand
is linear in prices. \Cref{sec:nonlinear} extends the analysis to
nonlinear demand.}

\vspace{1mm}
\begin{assumption}[Linear demand system]
\label{assump:linear-demand}
For each product $i\in\{1,\ldots,N\}$, the expected demand
function $F_i:\R^N\to\R$ is given by
\begin{equation}
F_i(\bp)
=
a_i+\sum_{j=1}^N b_{ij}p_j ,
\label{eq:demand-i}
\end{equation}
where the intercept $a_i$ represents the baseline demand of product $i$
when all prices are zero, and the slope $b_{ij}$ is the
\emph{price sensitivity of the demand for product $i$}
with respect to the price of product $j$.
In particular, $b_{ii}$ captures the own-price effect,
while $b_{ij}$ for $j\ne i$ captures cross-price effects.
\end{assumption}

\vspace{1mm}
Let $\ba=(a_1,\ldots,a_N)^\top$ denote the vector of intercepts, or
baseline demands, and let
$\bB=(b_{ij})_{i,j=1}^N$ denote the
\emph{price sensitivity matrix}.
Define the vector-valued demand function
$\bm{F}:\R^N\to\R^N$ by
$
\bm{F}(\bp)
=
\bigl(F_1(\bp),\ldots,F_N(\bp)\bigr)^\top .
$
Then, the expected demand formulation in \eqref{eq:demand-i} can be written compactly as
\[
\bm{F}(\bp)=\ba+\bB\bp.
\]

\noindent
The instantaneous revenue obtained at price vector
$\bp$ is
$R(\bp)
=
\sum_{i=1}^N p_i F_i(\bp)$.
This revenue admits the
quadratic representation
\[
R(\bp)
=
\ba^\top\bp+\bp^\top\bB\bp .
\]
Because $\epsilon_{i}$ has zero mean and is
independent of $\bm{p}$,
the firm's problem is to choose the price $\bm{p}$ to maximize the expected total revenue $ R(\bm{p})$. We study the design of pricing policies that maximize this total expected revenue.


\subsection{Additional Assumptions}

Motivated by economic and empirical considerations,
we impose several structural assumptions on the
price sensitivity matrix $\bB$.
These assumptions ensure that the revenue maximization
problem $\max_{\bp} R(\bp)$ admits a unique optimizer
and that the resulting analysis is economically
meaningful and mathematically tractable.
{We allow asymmetric cross-price effects, which implies that the matrix $B$ can be asymmetric. Throughout, we write that
\[
\bB_{\mathrm{s}}:=\frac{\bB+\bB^\top}{2},\qquad
\bD:=\mathrm{diag}(-b_{11},\ldots,-b_{NN}).
\]
The matrix $\bB_{\mathrm{s}}$ now becomes symmetric by definition. Since
$\bp^\top\bB\bp=\bp^\top\bB_{\mathrm{s}}\bp$, it is standard to use
$\bB_{\mathrm{s}}$ for aggregate-revenue curvature, while retaining
$\bB$ in individual demand functions and equilibrium conditions.
We impose the following assumptions.}

\vspace{1mm}
\begin{assumption}[Negative own-price effects]
\label{assump:own-price}
For each product $i\in\{1,\ldots,N\}$,
$
b_{ii}<0 .
$
\end{assumption}

\vspace{1mm}
Assumption~\ref{assump:own-price} states that the demand
for each product is strictly decreasing in its own
price, holding all other prices fixed.
Equivalently, we have
\[
\frac{\partial F_i(\bp)}{\partial p_i}=b_{ii}<0 ,
\]
which is the standard monotonicity condition in
revenue management and demand modeling. 

Before introducing further restrictions on $\bB$,
it is useful to interpret the economic meaning of
cross-price effects.
For distinct products $i\ne j$, the sign of $b_{ij}$
determines {the direction of the cross-price effect from product $j$
to the demand for product $i$}:
\begin{itemize}
\item
If $b_{ij}>0$, an increase in the price of product $j$
raises the demand for product $i$, {representing a substitution
effect in this direction}.
\item
If $b_{ij}<0$, an increase in the price of product $j$
reduces the demand for product $i$, {representing a complementarity
effect in this direction}.
\item
If $b_{ij}=0$, the demand for product $i$ is
independent of the price of product $j$.
\end{itemize}

{\vspace{1mm}
\noindent
Note that the effects $b_{ij}$ and $b_{ji}$ need not coincide, and may have different signs.}
Our next condition bounds the magnitude of
cross-price effects relative to own-price effects.

\vspace{1mm}
\begin{assumption}[{Two-sided strict diagonal dominance}]
\label{assump:diagdom}
There exists a scalar $\mu\in[0,1)$ such that, for each
$i\in\{1,\ldots,N\}$, we have
\begin{equation}
{\max\left\{\sum_{j\ne i}|b_{ij}|,
             \sum_{j\ne i}|b_{ji}|\right\}
\leq\mu\,|b_{ii}|.
\label{eq:diagdom}}
\end{equation}
\end{assumption}

{\vspace{1mm}
Assumption~\ref{assump:diagdom} requires that, for every
product $i$, the magnitude of its own-price effect
$|b_{ii}|$ strictly dominates the total magnitude of all
cross-price effects in both row and column $i$,
$\sum_{j\ne i}|b_{ij}|$ and $\sum_{j\ne i}|b_{ji}|$.
The parameter $\mu\in[0,1)$ quantifies the overall strength
of cross-product interactions.
When $\mu=0$, all cross-price effects vanish and products
are independent.
When $\mu$ is close to~$1$, cross-price effects are nearly
as strong as own-price effects, corresponding to a highly
interconnected market.}

From a matrix perspective, condition
\eqref{eq:diagdom} implies that the matrix $-\bB$ is
\emph{strictly diagonally dominant} with positive diagonal
entries.
Recall that a matrix $\mathbf{M}=(m_{ij})$ is said to be
strictly diagonally dominant if
$
|m_{ii}|>\sum_{j\ne i}|m_{ij}|
\text{ for all } i.
$
Since $b_{ii}<0$ by Assumption~\ref{assump:own-price}, we
have
\[
(-\bB)_{ii}=-b_{ii}=|b_{ii}|>0 .
\]
Moreover, because $\mu<1$, the inequality \eqref{eq:diagdom} implies
\[
|b_{ii}|
>
\mu |b_{ii}|
\ge
\sum_{j\ne i}|b_{ij}|
=
\sum_{j\ne i}|(-\bB)_{ij}|,
\]
which is exactly the strict diagonal dominance condition
for the matrix $-\bB$.

{Assumption~\ref{assump:diagdom} is a sufficient condition for the
pricing problem to be well posed. Together with
Assumption~\ref{assump:own-price}, it implies that $\bB_{\mathrm{s}}$ is
negative definite (Lemma~\ref{lem:B-negdef} below). Hence, the centralized problem
(to be defined in equation \eqref{eq:central-problem} later) has a finite and unique maximizer for every
intercept vector $\ba$, and the decentralized pricing game admits a unique
Nash equilibrium (Theorem~\ref{thm:NE-characterization}).
The threshold $\mu<1$ cannot be relaxed. For every $\mu\ge 1$, the class of matrices satisfying
\eqref{eq:diagdom} contains instances for which the centralized problem is unbounded above and thus has no maximizer.

\vspace{1mm}
\begin{remark}[Role of $\mu$ in the analysis]
\label{rem:mu}
The parameter $\mu$ is the central quantity governing the
strength of cross-product interactions in the model.
All of our subsequent results, including the negative
definiteness of the {symmetric part $\bB_{\mathrm{s}}$}, the existence and
uniqueness of the Nash equilibrium, and the bounds on the
price of anarchy, are driven by the diagonal dominance
condition parameterized by $\mu$.
In particular, the price-of-anarchy bounds derived in
later sections will be stated explicitly as functions of
$\mu$, making transparent how increasing interaction
strength degrades system-level performance.
\end{remark}

\vspace{1mm}
We now state a fundamental implication of {the structural assumptions and show that
only the symmetric part enters the aggregate-revenue quadratic form:
$\bp^\top\bB\bp=\bp^\top\bB_{\mathrm{s}}\bp$}.

\begin{lemma}[{Negative definiteness of the symmetric part}]
\label{lem:B-negdef}
Under Assumptions~{\ref{assump:own-price}} and~\ref{assump:diagdom},
{$\bB_{\mathrm{s}}$ is symmetric negative definite. In particular, it satisfies}
\[
{(1-\mu)\bD\preceq-\bB_{\mathrm{s}}\preceq(1+\mu)\bD.}
\]
\end{lemma}

{The matrix $\bB_{\mathrm{s}}^{-1}$ is symmetric} negative definite.
{Consequently, $R(\bp)=\ba^\top\bp+\bp^\top\bB_{\mathrm{s}}\bp$ is
strongly concave and enjoys a unique optimal solution.}

\subsection{The Centralized Pricing Scheme}

We now discuss the centralized pricing scheme, which
chooses all prices jointly to maximize total revenue.
This provides a natural benchmark against which the
performance of alternative pricing schemes will be
evaluated.

The centralized pricing problem is the unconstrained
optimization problem
\begin{equation}
R^*
=
\max_{\bp\in\R^N} R(\bp).
\label{eq:central-problem}
\end{equation}
By Lemma~\ref{lem:B-negdef}, $R(\bp)$ is a
concave function, and standard optimization
methods can be used to achieve the optimal value $R^*$.
For this reason, we use $R^*$ as the benchmark revenue
associated with centralized pricing.

The centralized optimizer and the associated maximum
revenue admit closed-form expressions.

\begin{lemma}
\label{thm:central-opt}
Under Assumptions~{\ref{assump:own-price}} and~\ref{assump:diagdom}, problem \eqref{eq:central-problem} has a
unique maximizer given by
\begin{equation*}
{\bp^*
=
-\frac{1}{2}\bB_{\mathrm{s}}^{-1}\ba .
\label{eq:p-star}}
\end{equation*}
The corresponding optimal revenue is
\begin{equation*}
{R(\bp^*)
=
-\frac{1}{4}\ba^\top\bB_{\mathrm{s}}^{-1}\ba .
\label{eq:R-star}}
\end{equation*}
\end{lemma}

Whenever $\ba\ne\mathbf{0}$, the optimal revenue
$R(\bp^*)$ is strictly positive.
Indeed, since {$\bB_{\mathrm{s}}^{-1}$} is symmetric and negative
definite, we have
\[
{\ba^\top\bB_{\mathrm{s}}^{-1}\ba<0
\quad
\text{for all } \ba\ne\mathbf{0}.}
\]
It follows that
\[
{R(\bp^*)
=
-\frac{1}{4}\ba^\top\bB_{\mathrm{s}}^{-1}\ba
>
0 ,}
\]
so the firm achieves strictly positive revenue whenever
baseline demand is nonzero.

\section{A Decentralized Pricing Scheme}
\label{sec:DecentralizedPricing}

We now consider a decentralized scheme. Per our discussion in \Cref{sec:Intro}, we focus on the decentralized pricing scheme determined by the Nash equilibrium of the following pricing game.

\vspace{1mm}
\begin{definition}[Decentralized pricing game]
\label{def:pricing-game}
The \emph{decentralized pricing game} is a simultaneous-move game defined as follows.
\begin{itemize}
\item
\textbf{Players.}
There are $N$ players, indexed by
$i\in\{1,\ldots,N\}$, where player $i$ controls the price
of product~$i$.

\item
\textbf{Actions.}
Each player $i$ chooses a price $p_i\in\R$.

\item
\textbf{Payoffs.}
Given a price vector
$\bp=(p_1,\ldots,p_N)^\top$, the payoff of player $i$ is
the revenue generated by product $i$,
\begin{equation*}
u_i(\bp)
=
p_i F_i(\bp)
=
p_i\!\left(
a_i+\sum_{j=1}^N b_{ij}p_j
\right).
\label{eq:payoff-i}
\end{equation*}
\end{itemize}
\end{definition}
\vspace{1mm}

In the pricing game defined in
Definition~\ref{def:pricing-game}, each player maximizes
only the revenue of its own product rather than the
firm’s total revenue.
As a result, the optimal price of any given product
depends on the prices chosen by all other products, and
these interdependencies create strategic tension across
players.
The Nash equilibrium concept formalizes a situation in
which no product can increase its own revenue by
unilaterally changing its price.

\vspace{1mm}
\begin{definition}[Pure-strategy Nash equilibrium]
\label{def:Nash-equilibrium}
A price vector
$\bp^{\mathrm{NE}}
=
(p_1^{\mathrm{NE}},\ldots,p_N^{\mathrm{NE}})^\top\in\R^N$
is a \emph{pure-strategy Nash equilibrium} of the
decentralized pricing game if, for every
$i\in\{1,\ldots,N\}$ and every alternative price
$p_i'\in\R$,
\begin{equation*}
u_i(\bp^{\mathrm{NE}})
\ge
u_i(p_i',\bp^{\mathrm{NE}}_{-i}),
\label{eq:Nash-condition}
\end{equation*}
where
$\bp^{\mathrm{NE}}_{-i}
=
(p_1^{\mathrm{NE}},\ldots,
p_{i-1}^{\mathrm{NE}},
p_{i+1}^{\mathrm{NE}},\ldots,
p_N^{\mathrm{NE}})^\top$
denotes the vector of equilibrium prices of all players
other than~$i$.
The notation $(p_i',\bp^{\mathrm{NE}}_{-i})$ refers to the
price vector obtained by replacing the $i$th component
of $\bp^{\mathrm{NE}}$ with $p_i'$ while keeping all other
components fixed.
\end{definition}

\vspace{1mm}
In other words, at a Nash equilibrium each player’s price
is a best response to the prices chosen by all other
players, and no player can increase its own revenue by
unilaterally deviating from its equilibrium price. 

We now establish the existence of a Nash equilibrium for
the decentralized pricing game defined in
Definition~\ref{def:pricing-game}.
To this end, we first study the optimization problem
faced by a single player when the prices of all other
products are held fixed.

Fix a player $i\in\{1,\ldots,N\}$ and suppose that the
prices chosen by all other players are given by
$\bp_{-i}\in\R^{N-1}$.
Player $i$ then chooses a price $p_i\in\R$ to maximize
its own payoff, which we denote by
$u_i(p_i;\bp_{-i})$ to emphasize that $p_i$ is the
decision variable and $\bp_{-i}$ is treated as fixed. It is easy to see that the payoff function of player $i$ can be written as
\begin{equation}
u_i(p_i;\bp_{-i})
=
a_i p_i
+
b_{ii} p_i^2
+
p_i\sum_{j\ne i} b_{ij} p_j ,
\label{eq:ui-expanded}
\end{equation}
which is a quadratic function of $p_i$.
Since we assume $b_{ii} < 0$, the function
$p_i\mapsto u_i(p_i;\bp_{-i})$ is strictly concave in $p_i$ and admits a
unique maximizer $p_i^*(\bp_{-i})$, which is the
\emph{best response} of player $i$ to the price vector
$\bp_{-i}$. The best response satisfies the first-order condition
\begin{equation}
\frac{\partial u_i}{\partial p_i}
\bigl(p_i^*(\bp_{-i});\bp_{-i}\bigr)=0 .
\label{eq:FOC-player-i}
\end{equation}
Solving \eqref{eq:FOC-player-i} using
\eqref{eq:ui-expanded} yields
\begin{equation}
p_i^*(\bp_{-i})
=
\frac{-a_i-\sum_{j\ne i} b_{ij}p_j}{2b_{ii}} .
\label{eq:best-response-explicit}
\end{equation}
Thus, the best response of each player is an affine
function of the prices chosen by the other players, a
direct consequence of the linear demand system.

At a Nash equilibrium, every player must be playing a
best response to the prices of all other players.
This yields a system of linear equations characterizing
equilibrium prices.

\vspace{1mm}
\begin{proposition}[First-order conditions at a Nash equilibrium]
\label{prop:FOC-Nash}
A price vector $\bp^{\mathrm{NE}}\in\R^N$ is a Nash
equilibrium of the decentralized pricing game if and
only if, for every $i\in\{1,\ldots,N\}$,
\begin{equation}
a_i
+
2b_{ii}p_i^{\mathrm{NE}}
+
\sum_{j\ne i} b_{ij}p_j^{\mathrm{NE}}
=
0 .
\label{eq:FOC-NE-coord}
\end{equation}
\end{proposition}
\vspace{1mm}

It is convenient to write the system of first-order
conditions \eqref{eq:FOC-NE-coord} in matrix form.

\vspace{1mm}
\begin{definition}[Nash equilibrium matrix]
\label{def:A-NE}
Define the matrix $\bA^{\mathrm{NE}}\in\R^{N\times N}$ by
\begin{equation*}
A^{\mathrm{NE}}_{ij}
=
\begin{cases}
2b_{ii}, & \text{if } i=j,\\[2pt]
b_{ij}, & \text{if } i\ne j .
\end{cases}
\label{eq:A-NE-def}
\end{equation*}
\end{definition}
\vspace{1mm}

Equivalently, $\bA^{\mathrm{NE}}$ is obtained from $\bB$
by doubling its diagonal entries,
\[
\bA^{\mathrm{NE}}
=
\bB+\mathrm{diag}(b_{11},\ldots,b_{NN}),
\]
where $\mathrm{diag}(b_{11},\ldots,b_{NN})$ denotes the
diagonal matrix with $(i,i)$-entry $b_{ii}$, equaling $-\bD$.

Using Definition~\ref{def:A-NE}, the system
\eqref{eq:FOC-NE-coord} for $i=1,\ldots,N$ can be written
compactly as
\begin{equation*}
\bA^{\mathrm{NE}}\bp^{\mathrm{NE}}+\ba=\mathbf{0},
\label{eq:NE-system-v1}
\end{equation*}
or, equivalently,
\begin{equation}
\bA^{\mathrm{NE}}\bp^{\mathrm{NE}}=-\ba .
\label{eq:NE-system}
\end{equation}
To ensure that this linear system admits a unique
solution, we require $\bA^{\mathrm{NE}}$ to be
invertible, which is established next.

\vspace{1mm}
\begin{lemma}
\label{lem:A-negdef}
Under Assumptions~{\ref{assump:own-price}} and~\ref{assump:diagdom}, we have that
\[
{\frac{\bA^{\mathrm{NE}}+(\bA^{\mathrm{NE}})^\top}{2}
=\bB_{\mathrm{s}}-\bD\prec\mathbf0.}
\]
{Hence}, the matrix $\bA^{\mathrm{NE}}$ is {invertible.}
\end{lemma}

\vspace{1mm}
We are now ready to state the main result of this section.

\vspace{1mm}
\begin{theorem}[Existence, uniqueness, and characterization
of the Nash equilibrium]
\label{thm:NE-characterization}
Under Assumptions {\ref{assump:own-price}} and \ref{assump:diagdom}, the
decentralized pricing game admits a unique pure-strategy
Nash equilibrium
$\bp^{\mathrm{NE}}\in\R^N$.
Moreover, this equilibrium is given by
\begin{equation*}
\bp^{\mathrm{NE}}
=
-(\bA^{\mathrm{NE}})^{-1}\ba .
\label{eq:p-NE}
\end{equation*}
\end{theorem}

We emphasize that the complete-information formulation above
serves to characterize the outcome of decentralized pricing, not to
describe what each player knows. Computing the equilibrium in closed
form requires $\ba$ and $\bB$, but reaching it does not: as discussed
in \Cref{rem:learning} below, the equilibrium arises as the limit of learning
dynamics in which each player observes only its own revenue, which is exactly the scenario that motivates this paper.

\vspace{1mm}
\begin{remark}\label{rem:learning}
The existence and uniqueness of the Nash equilibrium established in
\Cref{thm:NE-characterization} is important not only from a static
equilibrium perspective, but also from an algorithmic one.
In practice, a decentralized pricing scheme is often implemented
through adaptive or learning-based procedures, in which each player
updates its price using only local payoff information.
A natural approach is for each player to update its price using its own
marginal revenue. The negative marginal-revenue map is strongly monotone
because its Jacobian has symmetric part
$-\tfrac12(\bA^{\mathrm{NE}}+(\bA^{\mathrm{NE}})^\top)=-\bB_{\mathrm{s}}+\bD\succ0$, which makes the resulting
decentralized pricing game a strongly monotone game. It is therefore well known that when
players independently apply algorithms such as online gradient descent, the induced joint action
sequence converges to the unique Nash equilibrium, both in the time–averaged sense and in the
last–iterate sense \citep{ba2025doubly, jordan2025adaptive}.
Thus, the Nash equilibrium characterized in
\Cref{thm:NE-characterization} naturally emerges as the convergence
point of decentralized learning dynamics.
\end{remark}

\vspace{1mm}

Combining \Cref{thm:central-opt} and
\Cref{thm:NE-characterization}, we obtain explicit
expressions for the centralized and decentralized price
vectors:
\[
{\bp^*
=
-\tfrac{1}{2}\bB_{\mathrm{s}}^{-1}\ba,
\qquad
\bp^{\mathrm{NE}}
=
-(\bA^{\mathrm{NE}})^{-1}\ba .}
\]
Both price vectors depend linearly on the intercept
vector $\ba$, but they are generated by different
matrices.
In general, $\bp^*\ne\bp^{\mathrm{NE}}$, reflecting the
inefficiency induced by decentralized pricing.
The next sections quantify this inefficiency through the
price of anarchy.

\section{Price of Anarchy and Tightness Analysis}
\label{sec:PoA-analysis}

In this section, we compare the centralized optimal
revenue $R(\bp^*)$ with the revenue attained at the
decentralized Nash equilibrium $R(\bp^{\mathrm{NE}})$. 

\subsection{Mathematical Expression of Price of Anarchy}

We start by noting that $R(\bp^*)$ and $R(\bp^{\mathrm{NE}})$ can be written as quadratic forms in
the intercept vector $\ba$.
This representation allows us to define and analyze the
\emph{price of anarchy}, which measures the worst-case
ratio between centralized and decentralized revenues
over all admissible instances.

Recall from Theorem~\ref{thm:central-opt} that the
centralized optimal prices and revenue are
\[
{\bp^*
=
-\frac{1}{2}\bB_{\mathrm{s}}^{-1}\ba,
\qquad
R(\bp^*)
=
-\frac{1}{4}\ba^\top\bB_{\mathrm{s}}^{-1}\ba .}
\]
From Theorem~\ref{thm:NE-characterization}, the Nash
equilibrium prices are
\[
\bp^{\mathrm{NE}}
=
-(\bA^{\mathrm{NE}})^{-1}\ba .
\]
The corresponding equilibrium revenue
$R(\bp^{\mathrm{NE}})$ admits an analogous quadratic
representation.
Whereas $R(\bp^*)$ is proportional to
{$\ba^\top\bB_{\mathrm{s}}^{-1}\ba$}, we will show that
$R(\bp^{\mathrm{NE}})$ can be written as
$\ba^\top\bK\ba$ for a suitable matrix $\bK$, which we
introduce next.

{\vspace{1mm}
\begin{definition}[Nash revenue matrix]
\label{def:K-matrix}
Define the matrix $\bK\in\R^{N\times N}$ by
\begin{equation*}
\bK:=-4(\bA^{\mathrm{NE}})^{-\top}\bD(\bA^{\mathrm{NE}})^{-1}.
\label{eq:K-def}
\end{equation*}
This matrix is symmetric negative definite.
\end{definition}}
\vspace{1mm}

Note that the normalization factor of $4$ is chosen so that the
Nash equilibrium revenue admits the representation
\[
R(\bp^{\mathrm{NE}})
=
-\frac{1}{4}\ba^\top\bK\ba,
\]
which mirrors the centralized revenue formula
\[
{R(\bp^*)
=
-\frac{1}{4}\ba^\top\bB_{\mathrm{s}}^{-1}\ba .}
\]
This parallel form will simplify the subsequent
price-of-anarchy analysis.

We now formalize the notion of efficiency loss induced by
decentralized pricing.

\vspace{1mm}
\begin{definition}[Price of anarchy]
\label{def:PoA}
For any nonzero intercept vector $\ba\ne\mathbf{0}$, the
\emph{price of anarchy} is defined as
\begin{equation*}
\mathrm{PoA}(\ba)
:=
\frac{R(\bp^{\mathrm{NE}})}{R(\bp^*)}.
\label{eq:PoA-def}
\end{equation*}
\end{definition}

\vspace{1mm}
The price of anarchy measures the fraction of the
centralized optimal revenue achieved at the Nash
equilibrium.
When $\mathrm{PoA}(\ba)=1$, decentralization is fully
efficient.
When $\mathrm{PoA}(\ba)<1$, decentralized pricing leads
to a revenue loss equal to $1-\mathrm{PoA}(\ba)$.
Since $\bp^*$ maximizes revenue, we always have
$R(\bp^{\mathrm{NE}})\le R(\bp^*)$, and therefore
$\mathrm{PoA}(\ba)\le 1$.

Using Theorems \ref{thm:NE-characterization} and \ref{thm:central-opt}, the price of anarchy can be
written as a ratio of two quadratic forms in the
intercept vector $\ba$.

\vspace{1mm}
\begin{proposition}[Price of anarchy expression]
\label{prop:PoA-ratio}
For any nonzero $\ba\in\R^N$, we have $R(\bp^{\mathrm{NE}})
=
-\frac{1}{4}\ba^\top\bK\ba$, which leads to
\begin{equation}
{\mathrm{PoA}(\ba)
=
\frac{\ba^\top\bK\ba}{\ba^\top\bB_{\mathrm{s}}^{-1}\ba}.
\label{eq:PoA-ratio}}
\end{equation}
\end{proposition}

\vspace{1mm}
To facilitate our analysis of $\mathrm{PoA}(\ba)$, we next show that it can be written as a Rayleigh quotient of a symmetric positive definite matrix, and
its extreme values over all intercept vectors are given
by the eigenvalues of this matrix.

\subsection{Spectral Representation of the Price of Anarchy}
\label{subsec:PoA-spectral}


The matrices {$\bB_{\mathrm{s}}^{-1}$} and $\bK$ appearing in
\eqref{eq:PoA-ratio} are associated with negative
definite matrices, since {$\bB_{\mathrm{s}}$} is negative definite.
For spectral analysis it is more convenient to work with
positive definite objects, which motivates the
following definitions.

\vspace{1mm}
\begin{definition}[Positive definite revenue matrices]
\label{def:Ltilde-Ktilde}
Define $\tilde{\bL},\tilde{\bK}\in\R^{N\times N}$ by
\begin{equation*}
{\tilde{\bL}:=-\bB_{\mathrm{s}}^{-1},
\qquad
\tilde{\bK}:=-\bK .}
\end{equation*}
\end{definition}

\vspace{1mm}
By Lemma~\ref{lem:B-negdef}, $\tilde{\bL}$ is symmetric
and positive definite.
Using Proposition~\ref{prop:PoA-ratio}, we can rewrite
the price of anarchy as
\begin{equation}
\mathrm{PoA}(\ba)
=
\frac{\ba^\top\tilde{\bK}\ba}{\ba^\top\tilde{\bL}\ba},
\qquad
\ba\ne\mathbf{0}.
\label{eq:PoA-tilde}
\end{equation}

\vspace{1mm}
\noindent
The key step is to convert the generalized Rayleigh
quotient in \eqref{eq:PoA-tilde} into a standard one.
The expression in \eqref{eq:PoA-tilde} involves two
matrices, $\tilde{\bK}$ and $\tilde{\bL}$, and therefore
defines a \emph{generalized Rayleigh quotient}.
To apply the classical spectral characterization, in
which extrema are given by eigenvalues of a single
matrix, the denominator must take the form
$\mathbf{x}^\top\mathbf{x}$.

This is achieved by the change of variables
$\mathbf{x}=\tilde{\bL}^{1/2}\ba$.
Under this transformation, the denominator becomes
$\mathbf{x}^\top\mathbf{x}$, while the numerator is
mapped to $\mathbf{x}^\top\bM\mathbf{x}$ for a suitable
matrix $\bM$.
This reduces the generalized eigenvalue problem to a
standard eigenvalue problem.

\vspace{1mm}
\begin{definition}[Price-of-anarchy matrix]
\label{def:M-matrix}
Define the matrix $\bM\in\R^{N\times N}$ by
\begin{equation*}
\bM
:=
\tilde{\bL}^{-1/2}\,
\tilde{\bK}\,
\tilde{\bL}^{-1/2},
\label{eq:M-def}
\end{equation*}
where $\tilde{\bL}^{-1/2}$ denotes the unique symmetric
positive definite square root of $\tilde{\bL}^{-1}$.
\end{definition}
\vspace{1mm}

We briefly justify the existence of the matrix square
root $\tilde{\bL}^{-1/2}$.
For any symmetric positive definite matrix
$\mathbf{P}$, there exists a unique symmetric positive
definite matrix $\mathbf{P}^{1/2}$ such that
$\mathbf{P}^{1/2}\mathbf{P}^{1/2}=\mathbf{P}$.
This matrix is called the \emph{symmetric positive
definite square root} of $\mathbf{P}$.

One way to construct $\mathbf{P}^{1/2}$ is through the
eigen-decomposition of $\mathbf{P}$.
If
$\mathbf{P}=\mathbf{Q}\mathbf{\Lambda}\mathbf{Q}^\top$,
where $\mathbf{Q}$ is orthogonal and
$\mathbf{\Lambda}=\mathrm{diag}(\lambda_1,\ldots,
\lambda_N)$ with $\lambda_i>0$, then
\[
\mathbf{P}^{1/2}
=
\mathbf{Q}\mathbf{\Lambda}^{1/2}\mathbf{Q}^\top,
\qquad
\mathbf{\Lambda}^{1/2}
=
\mathrm{diag}(\sqrt{\lambda_1},\ldots,\sqrt{\lambda_N}).
\]

Since $\tilde{\bL}$ is symmetric positive definite by
Lemma~\ref{lem:B-negdef}, its inverse $\tilde{\bL}^{-1}$
is also symmetric positive definite.
Therefore the symmetric positive definite square root
$\tilde{\bL}^{-1/2}$ exists, is unique, and is itself
symmetric.

Before stating our main theorem in this subsection, we introduce several
auxiliary matrices that simplify the spectral analysis.
The {matrix $-\bB_{\mathrm{s}}$ is symmetric positive definite, whereas
$-\bA^{\mathrm{NE}}$ need not be symmetric}.

{\begin{definition}[Revenue curvature and equilibrium matrices]
\label{def:H-G-matrices}
\begin{equation*}
\bH:=-\bB_{\mathrm{s}},\qquad
\bG:=-\bA^{\mathrm{NE}}=\bD-\bB.
\label{eq:H-G-def}
\end{equation*}
Here $\bH\succ0$, $\bG$ is invertible, and
$(\bG+\bG^\top)/2=\bH+\bD$.
\end{definition}}

\vspace{1mm}
We next introduce an auxiliary matrix that will play a
central role in the spectral representation.

\vspace{1mm}
\begin{definition}[Auxiliary matrix $\bY$]
\label{def:Y-matrix}
Define $\bY\in\R^{N\times N}$ by
\begin{equation*}
\bY
:=
\bH^{1/2}\bG^{-1}\bH^{1/2},
\label{eq:Y-def}
\end{equation*}
where $\bH^{1/2}$ denotes the symmetric positive definite
square root of $\bH$.
\end{definition}

\vspace{1mm}
The following lemma establishes the link between the
price-of-anarchy matrix $\bM$ and the auxiliary matrix
$\bY$.

{\vspace{1mm}
\begin{lemma}[Eigenvalue characterization]
\label{lem:M-matrix}
Under Assumptions~\ref{assump:own-price} and~\ref{assump:diagdom}, the
price-of-anarchy matrix $\bM$ satisfies
\begin{align*}
\bM&=4\bH^{1/2}\bG^{-\top}\bD\bG^{-1}\bH^{1/2}\succ\mathbf0,
\\
\bM&=2(\bY+\bY^\top)-4\bY^\top\bY
=\bI-(\bI-2\bY)^\top(\bI-2\bY)\preceq\bI.
\end{align*}
Thus its eigenvalues belong to $(0,1]$.
\end{lemma}}

\vspace{1mm}
We are now ready to state the main result of this
subsection.

{\begin{theorem}[Spectral representation of the price of anarchy]
\label{thm:PoA-spectral}
Under Assumptions~\ref{assump:own-price} and~\ref{assump:diagdom}, for any
nonzero $\ba\in\R^N$, define
$\mathbf{x}:=\tilde{\bL}^{1/2}\ba$.
Then
\begin{equation*}
\mathrm{PoA}(\ba)
=
\frac{\mathbf{x}^\top\bM\mathbf{x}}
{\mathbf{x}^\top\mathbf{x}} .
\label{eq:PoA-Rayleigh}
\end{equation*}
Moreover, the best-case and worst-case prices of anarchy
over all nonzero $\ba$ satisfy
\begin{equation*}
\inf_{\ba\ne\mathbf{0}} \, \mathrm{PoA}(\ba)
=
\lambda_{\min}(\bM),
\qquad
\sup_{\ba\ne\mathbf{0}} \, \mathrm{PoA}(\ba)
=
\lambda_{\max}(\bM),
\label{eq:PoA-extrema}
\end{equation*}
where $\lambda_{\min}(\bM)$ and $\lambda_{\max}(\bM)$
denote the smallest and largest eigenvalues of $\bM$,
respectively.

\end{theorem}}

\vspace{1mm}
By Lemma~\ref{lem:M-matrix}, all eigenvalues of $\bM$ lie
in $(0,1]$, which implies
$0<\mathrm{PoA}(\ba)\le 1$ for all $\ba\ne\mathbf{0}$.
The upper bound reflects the fact that decentralized
pricing cannot outperform the centralized optimum.
The lower bound will be sharpened in the next subsection,
where we derive an explicit worst-case guarantee on the
efficiency of decentralization.

\subsection{A Worst-case Bound}
\label{sec:PoA-mu}

In this subsection, we derive an explicit lower bound for the price of anarchy in terms of the diagonal-dominance parameter $\mu$ from
Assumption~\ref{assump:diagdom}.

Recall from Theorem~\ref{thm:PoA-spectral} that, for a
fixed demand matrix $\bB$,
\[
\mathrm{PoA}_{\min}(\bB)
:=
\inf_{\ba\ne\mathbf{0}} \, \mathrm{PoA}(\ba)
=
\lambda_{\min}(\bM),
\]
where $\bM$ is the symmetric positive definite matrix
defined in Definition~\ref{def:M-matrix}.
Our goal is to bound $\lambda_{\min}(\bM)$ from below
using only the parameter~$\mu$.

{The following lemma controls the normalized directed interactions and
will be used to bound the revenue loss directly.}

{\vspace{1mm}
\begin{lemma}[Normalized interaction bound]
\label{lem:interaction-norm}
Under Assumptions~\ref{assump:own-price} and~\ref{assump:diagdom}, let
$\mathbf C:=\bB+\bD$ be the off-diagonal part of $\bB$ and set
$\mathbf T:=\bD^{-1/2}\mathbf C\bD^{-1/2}$. Then, we have that
\[
\|\mathbf T\|_2\leq\mu,
\qquad
\mathbf C\bH^{-1}\mathbf C^\top
\preceq\frac{\mu^2}{1-\mu}\bD,
\]
where $\|\cdot\|_2$ denotes the spectral norm, that is, the largest
singular value.
\end{lemma}}

We now state the main result of this subsection, which
provides an explicit lower bound on the price
of anarchy in terms of the diagonal-dominance parameter
$\mu$.

\vspace{1mm}
\begin{theorem}[Price of anarchy bound under diagonal dominance]
\label{thm:PoA-mu-bound}
Suppose that Assumptions~{\ref{assump:own-price}} and~\ref{assump:diagdom} hold with
diagonal-dominance parameter $\mu\in[0,1)$. Then, the
worst-case price of anarchy satisfies
\begin{equation}
\mathrm{PoA}_{\min}(\bB)
:=
\inf_{\ba\ne\mathbf{0}} \, \mathrm{PoA}(\ba)
=
\lambda_{\min}(\bM)
\ge
\frac{4(1-\mu)}{(2-\mu)^2}.
\label{eq:PoA-mu-bound}
\end{equation}
\end{theorem}
\vspace{1mm}

The bound in \eqref{eq:PoA-mu-bound} has several natural
properties.
When $\mu=0$, it reduces to
$4/2^2=1$, which implies
$\mathrm{PoA}(\ba)=1$ for all $\ba\ne\mathbf{0}$.
This is consistent with the case of independent
products, where $b_{ij}=0$ for $i\ne j$ and decentralized
pricing is fully efficient. Define
$f(\mu)=4(1-\mu)/(2-\mu)^2$. A direct calculation on the derivative of $f$ with respect to $\mu$ also shows that $f$ is concave and is strictly (slowly) decreasing on $[0,1)$. Hence, when $\mu$ is not too large, decentralized pricing achieves performance close to that of centralized pricing. Finally, as $\mu\to1^-$, the bound converges to zero, indicating
that when cross-effects are comparable to own-price
effects, decentralized pricing can become arbitrarily
inefficient in the worst case.

\vspace{1mm}
\begin{remark}[Numerical values of the bound]
\label{rem:PoA-bound-numerical}
The values
of the bound $4(1-\mu)/(2-\mu)^2$ are plotted in \Cref{fig:ratio}.
For instance, when $\mu$ is at most 0.5, 
the Nash
equilibrium revenue is guaranteed to be at least
$88.9\%$ of the centralized optimum.
\end{remark}

\begin{figure}[h]
    \centering
    \includegraphics[width=0.5\linewidth]{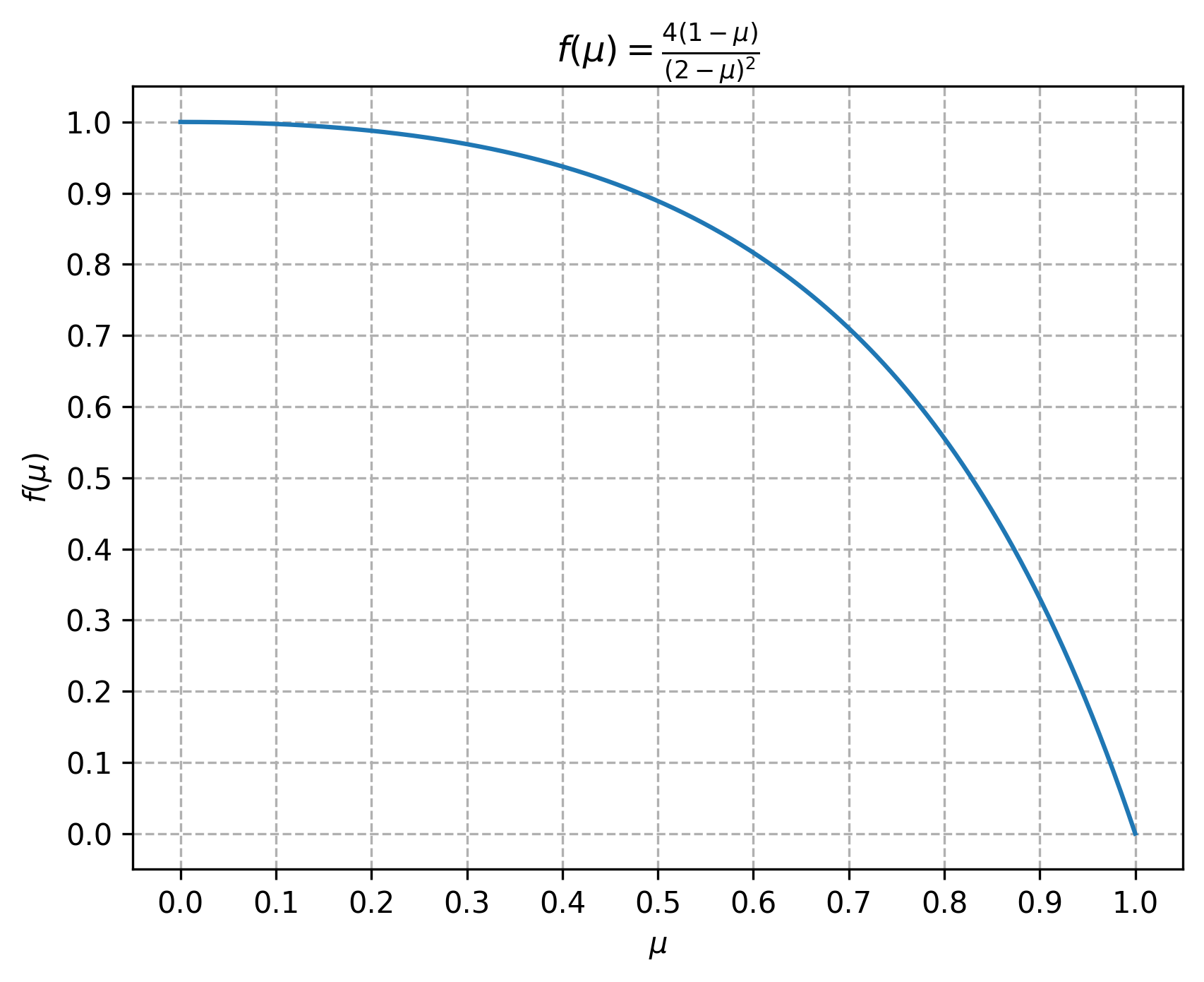}
    \caption{The values of the bound $4(1-\mu)/(2-\mu)^2$ for different $\mu$.}
    \label{fig:ratio}
\end{figure}

\subsection{Tightness Analysis}
\label{sec:Tightness}

In this section, we examine the tightness of the
price-of-anarchy bound established in
Theorem~\ref{thm:PoA-mu-bound}.
We show that the bound is attained by an important and
highly symmetric demand system, which implies that the
lower bound is sharp.

\vspace{1mm}
\begin{definition}[Fully symmetric exchangeable model]
\label{def:symmetric-model}
The fully symmetric exchangeable model with $N$ products
and interaction parameter $\rho\ge 0$ is defined by
\begin{equation*}
b_{ii}=-1
\quad\text{for all } i,
\qquad
b_{ij}=\rho
\quad\text{for all } i\ne j .
\end{equation*}
Equivalently,
\begin{equation}
\bB
=
-\bI+\rho(\mathbf{1}\mathbf{1}^\top-\bI)
=
-(1+\rho)\bI+\rho\,\mathbf{1}\mathbf{1}^\top,
\label{eq:symmetric-B-matrix}
\end{equation}
where $\mathbf{1}=(1,\ldots,1)^\top\in\R^N$ denotes the
all-ones vector.
\end{definition}

\vspace{1mm}
In this model, all products have identical own-price
sensitivities, normalized to $b_{ii}=-1$ without loss of
generality, since any symmetric system with equal
own-price effects can be rescaled to this form.
All cross-price effects are identical and positive,
$b_{ij}=\rho$ for $i\ne j$, so the products are mutual
substitutes: increasing the price of any one product
raises the demand for every other product.

A direct calculation shows that, for the fully symmetric
exchangeable model with parameters $N$ and $\rho$, the
diagonal-dominance parameter is
\begin{equation*}
\mu=(N-1)\rho .
\label{eq:symmetric-mu}
\end{equation*}
Hence, Assumption~\ref{assump:diagdom} holds if and only if
$\rho<1/(N-1)$.
As the number of products $N$ grows, this restriction
becomes more stringent, reflecting the fact that even
small pairwise interactions can accumulate across many
products and violate diagonal dominance.

The special structure of the symmetric model allows
closed-form spectral calculations.
In particular, $\bB$ is a rank-one perturbation of a
scaled identity matrix, which yields explicit formulas
for its eigenvalues and inverse.

\vspace{1mm}
\begin{lemma}[Spectrum of the symmetric demand matrix]
\label{lem:symmetric-B-eigenvalues}
In the fully symmetric exchangeable model, the matrix
$\bB$ has eigenvalue
$
\lambda_1=-1+(N-1)\rho
$
with eigenvector $\mathbf{1}$, and eigenvalue
$
\lambda_2=-1-\rho
$
with multiplicity $N-1$, corresponding to any vector
orthogonal to $\mathbf{1}$.
Moreover,
\begin{equation*}
\bB^{-1}
=
-\frac{1}{1+\rho}\bI
-
\frac{\rho}{(1+\rho)\bigl(1-(N-1)\rho\bigr)}
\,\mathbf{1}\mathbf{1}^\top .
\label{eq:symmetric-B-inverse}
\end{equation*}
\end{lemma}
\vspace{1mm}

We now compute the centralized and Nash equilibrium
revenues for the symmetric model under a symmetric
intercept vector.
Because $\bB$ is invariant under any simultaneous
permutation of rows and columns, any solution of the
first-order conditions with a symmetric intercept
vector $\ba=a\mathbf{1}$ must itself be symmetric.
Consequently, both the centralized and Nash equilibrium
price vectors lie in the one-dimensional subspace
spanned by $\mathbf{1}$.

\vspace{1mm}
\begin{lemma}[Centralized optimum in the symmetric model]
\label{prop:symmetric-optimal}
In the fully symmetric exchangeable model with intercept
vector $\ba=a\mathbf{1}$ and diagonal-dominance parameter
$\mu=(N-1)\rho$, the centralized optimal prices are
$\bp^*=p^*\mathbf{1}$ with
\begin{equation*}
p^*=\frac{a}{2(1-\mu)} ,
\label{eq:symmetric-p-star}
\end{equation*}
and the corresponding optimal revenue is
\begin{equation*}
R(\bp^*)
=
\frac{N a^2}{4(1-\mu)} .
\label{eq:symmetric-R-star}
\end{equation*}
\end{lemma}
\vspace{1mm}

We obtain analogous closed-form expressions for the
Nash equilibrium.

\vspace{1mm}
\begin{lemma}[Nash equilibrium in the symmetric model]
\label{prop:symmetric-nash}
In the fully symmetric exchangeable model with intercept
vector $\ba=a\mathbf{1}$ and diagonal-dominance parameter
$\mu=(N-1)\rho$, the Nash equilibrium prices are
$\bp^{\mathrm{NE}}=p^{\mathrm{NE}}\mathbf{1}$ with
\begin{equation*}
p^{\mathrm{NE}}=\frac{a}{2-\mu},
\label{eq:symmetric-p-NE}
\end{equation*}
and the corresponding Nash equilibrium revenue is
\begin{equation*}
R(\bp^{\mathrm{NE}})
=
\frac{N a^2}{(2-\mu)^2}.
\label{eq:symmetric-R-NE}
\end{equation*}
\end{lemma}
\vspace{1mm}

Combining Lemmas~\ref{prop:symmetric-optimal} and
\ref{prop:symmetric-nash} yields the following result.

\vspace{1mm}
\begin{theorem}[Tightness of the price-of-anarchy bound]
\label{thm:PoA-tight}
In the fully symmetric exchangeable model with intercept
vector $\ba=a\mathbf{1}$ and diagonal-dominance parameter
$\mu=(N-1)\rho<1$, we have
\begin{equation*}
\mathrm{PoA}(\ba)
=
\frac{R(\bp^{\mathrm{NE}})}{R(\bp^*)}
=
\frac{4(1-\mu)}{(2-\mu)^2}.
\label{eq:PoA-symmetric}
\end{equation*}

\vspace{1mm}
\noindent
This value coincides with the lower bound in \Cref{thm:PoA-mu-bound}, showing that the bound is
tight.
\end{theorem}
\vspace{1mm}

{The fully symmetric exchangeable model therefore
represents a worst-case instance for decentralized
pricing.
Here all products are identical and interact
symmetrically, which maximizes the inefficiency induced
by uncoordinated price setting.
Moreover, the intercept vector $\ba=a\mathbf{1}$ aligns with the
uniform interaction mode of this example and attains the
worst-case price of anarchy.}

\section{An Exact Characterization of PoA via Spectral Bound}
\label{sec:spectral}

{In \Cref{sec:PoA-analysis} we showed that the worst-case price of
anarchy equals the smallest eigenvalue of
$\bM=4\bH^{1/2}\bG^{-\top}\bD\bG^{-1}\bH^{1/2}$,
where $\bH=-\bB_{\mathrm{s}}$ and $\bG=\bD-\bB$, and derived a universal
lower bound in terms of the diagonal-dominance parameter $\mu$.
We further established in \Cref{sec:Tightness} that this bound is tight
for the fully symmetric exchangeable model, in which all products
interact with equal strength. When interaction strengths are
heterogeneous, however, this bound can be conservative.}

{The goal of this section is to provide an \emph{exact} spectral
characterization of the worst-case price of anarchy for a fixed,
possibly asymmetric demand matrix. The exact formula uses a symmetric
positive semidefinite matrix of normalized revenue externalities.
A scalar spectral-norm parameter gives a convenient lower bound.}

\subsection{The Normalized Interaction Matrix}
\label{subsec:normalized-interaction}

{We normalize cross-price interactions by the own-price
magnitudes $-b_{ii}>0$.}

\vspace{1mm}
\begin{definition}[Normalized interaction matrix]
\label{def:M-norm}
The \emph{normalized interaction matrix}
$\bM_{\mathrm{norm}}\in\R^{N\times N}$ is defined by
\begin{equation*}
(\bM_{\mathrm{norm}})_{ij}
=
\begin{cases}
\dfrac{b_{ij}}{{\sqrt{(-b_{ii})(-b_{jj})}}},
& i\ne j,\\[6pt]
0, & i=j .
\end{cases}
\label{eq:M-norm-entries}
\end{equation*}
\end{definition}

{\vspace{1mm}
By construction, $\bM_{\mathrm{norm}}$ has zero diagonal and
possibly asymmetric off-diagonal entries. For $i\neq j$, the entry
\[
(M_{\mathrm{norm}})_{ij}=\frac{b_{ij}}{\sqrt{(-b_{ii})(-b_{jj})}}
\]
is a \emph{dimensionless} measure of the strength of the cross-price interaction between products
$i$ and $j$, normalized by the geometric mean of their own-price sensitivities.
Thus, $|(M_{\mathrm{norm}})_{ij}|$ is large when the cross effect $b_{ij}$ is large relative to the two own effects,
and small when cross effects are weak compared to own effects.
The sign of $(M_{\mathrm{norm}})_{ij}$ retains the economic direction of interaction (e.g., positive for
substitution and negative for complementarity in the corresponding direction).
Equivalently, $M_{\mathrm{norm}}=D^{-1/2}B_{\mathrm{off}}D^{-1/2}$ rescales each product so that own-price
effects are comparable across products; the resulting matrix isolates \emph{relative} cross-product coupling
while accounting for heterogeneous own-price magnitudes $\{-b_{ii}\}$.
If $b_{ii}=b_{11}$ for all $i$, then $M_{\mathrm{norm}}=B_{\mathrm{off}}/(-b_{11})$.}

Let
$
\bD:={\mathrm{diag}(-b_{11},\ldots,-b_{NN})}
$
denote the diagonal matrix of own-price effect
magnitudes, and let $\bB_{\mathrm{off}}$ be the matrix
with entries $b_{ij}$ for $i\ne j$ and zeros on the
diagonal.
Then, we have
\begin{equation*}
\bM_{\mathrm{norm}}
=
\bD^{-1/2}\,\bB_{\mathrm{off}}\,\bD^{-1/2}.
\label{eq:M-norm-matrix}
\end{equation*}
{Define its symmetric part by
\begin{equation*}
\mathbf S:=\frac{\bM_{\mathrm{norm}}+\bM_{\mathrm{norm}}^\top}{2}
=\bD^{-1/2}(\bB_{\mathrm{s}}+\bD)\bD^{-1/2}.
\end{equation*}}
{Then, we have $\bH=\bD^{1/2}(\bI-\mathbf S)\bD^{1/2}$, and
$\bI-\mathbf S\succ0$. We now introduce the following parameter.}

{\vspace{1mm}
\begin{definition}[Spectral interaction parameter]
\label{def:mu-spectral}
The \emph{spectral interaction parameter} is defined as
\begin{equation*}
\mu_{\mathrm{spectral}}:=\|\bM_{\mathrm{norm}}\|_2
=\sigma_{\max}(\bM_{\mathrm{norm}}),
\label{eq:mu-spectral-def}
\end{equation*}
where $\sigma_{\max}(\cdot)$ denotes the largest singular value.
\end{definition}}
\vspace{1mm}

{The next lemma summarizes basic properties of
$\bM_{\mathrm{norm}}$ and $\mu_{\mathrm{spectral}}$.}

{\vspace{1mm}
\begin{lemma}[Properties of $\bM_{\mathrm{norm}}$]
\label{lem:M-norm-properties}
Under Assumptions~\ref{assump:own-price} and~\ref{assump:diagdom}, we have that
\[
0\leq\mu_{\mathrm{spectral}}
\leq\max_{1\leq i\leq N}\mu_i\leq\mu<1,
\qquad
\mu_i:=\frac{1}{-b_{ii}}\max\left\{
\sum_{j\ne i}|b_{ij}|,\sum_{j\ne i}|b_{ji}|\right\}.
\]
All eigenvalues of the symmetric matrix $\mathbf S$ belong to
$[-\mu_{\mathrm{spectral}},\mu_{\mathrm{spectral}}]$.
\end{lemma}}

{\vspace{1mm}
The smallest admissible choice of $\mu$ in
Assumption~\ref{assump:diagdom} is $\max_i\mu_i$.
Unlike a spectral radius, $\mu_{\mathrm{spectral}}$ controls singular
values even when $\bM_{\mathrm{norm}}$ is not symmetric.}

\subsection{Exact Formula for the {Worst-case} Price of Anarchy}
\label{subsec:PoA-exact}

{We now characterize the worst-case price of anarchy using the full
directed interaction matrix. Define
\begin{equation}
\bQ:=\bM_{\mathrm{norm}}(\bI-\mathbf S)^{-1}              \bM_{\mathrm{norm}}^\top\succeq\mathbf0.
\label{eq:externality-matrix}
\end{equation}}
{This matrix is symmetric by construction. We have the following result.}

{\vspace{1mm}
\begin{lemma}[Revenue-gap representation]
\label{lem:revenue-gap}
Under Assumptions~\ref{assump:own-price} and~\ref{assump:diagdom}, let
$\mathbf x=\bp^{\mathrm{NE}}=\bG^{-1}\ba$ and
$\mathbf C=\bB+\bD$. Then, we have that
\begin{equation*}
R(\bp^*)-R(\mathbf x)
=\frac14\mathbf x^\top\mathbf C\bH^{-1}\mathbf C^\top\mathbf x.
\end{equation*}
For $\ba\ne\mathbf0$, with $\bz=\bD^{1/2}\mathbf x$, it holds that
\[
\mathrm{PoA}(\ba)
=\left(1+\frac{\bz^\top\bQ\bz}{4\bz^\top\bz}\right)^{-1}.
\]
\end{lemma}}

\vspace{1mm}
We are now ready to state the main result of this section.

{\vspace{1mm}
\begin{theorem}[Exact spectral formula for the price of anarchy]
\label{thm:PoA-spectral-exact}
Under Assumptions~\ref{assump:own-price} and~\ref{assump:diagdom}, the
worst-case price of anarchy satisfies
\begin{equation}
\mathrm{PoA}_{\min}(\bB)
=\left(1+\frac14\lambda_{\max}(\bQ)\right)^{-1}
=\frac{4}{4+\|\bM_{\mathrm{norm}}(\bI-\mathbf S)^{-1/2}\|_2^2}.
\label{eq:PoA-exact}
\end{equation}
The minimum is attained at $\ba^*=\bG\bD^{-1/2}\bv^*$, where $\bv^*$
is any nonzero eigenvector of $\bQ$ associated with
$\lambda_{\max}(\bQ)$. In particular, we have
\begin{equation}
\mathrm{PoA}_{\min}(\bB)
\geq\frac{4(1-\mu_{\mathrm{spectral}})}
{(2-\mu_{\mathrm{spectral}})^2}
\geq\frac{4(1-\mu)}{(2-\mu)^2}.
\label{eq:PoA-spectral-norm-bound}
\end{equation}
\end{theorem}}

{\vspace{1mm}
\Cref{thm:PoA-spectral-exact} refines \Cref{thm:PoA-mu-bound} by
identifying the exact worst-case value for each demand matrix.}
{For a matrix with $\bM_{\mathrm{norm}}=\bM_{\mathrm{norm}}^\top$,
\eqref{eq:externality-matrix} reduces algebraically to
$\bQ=\bM_{\mathrm{norm}}^2(\bI-\bM_{\mathrm{norm}})^{-1}$.
If its eigenvalues are $\lambda_1,\ldots,\lambda_N$, the exact formula
therefore reduces to
\[
\mathrm{PoA}_{\min}(\bB)
=\min_i\frac{4(1-\lambda_i)}{(2-\lambda_i)^2}.
\]
With $f(t)=4(1-t)/(2-t)^2$, we have that
\[
\mathrm{PoA}_{\min}=\min\{f(\lambda_{\min}),f(\lambda_{\max})\}.
\]
We attains the lower bound
$f(\mu_{\mathrm{spectral}})$ if and only if
$\lambda_{\max}=\mu_{\mathrm{spectral}}$, which recovers and strengthens \Cref{thm:PoA-mu-bound}.}

\subsection{Examples}
\label{subsec:spectral-examples}

In this subsection, we illustrate the spectral characterization of the price
of anarchy through two representative examples.

{\vspace{1mm}
\begin{example}[Uniform local connectivity]
\label{ex:uniform}
Suppose $b_{ij}\geq0$ for all $i\ne j$, all own-price effects
are equal, $b_{ii}=b_{11}$ for all $i$, and each row satisfies
$\sum_{j\ne i}b_{ij}/(-b_{ii})=\mu$. This includes the exchangeable model
in \cref{sec:Tightness}, but does not require identical pairwise
interactions. Then, we have
\[
\bM_{\mathrm{norm}}\mathbf1=\mu\mathbf1.
\]
By Assumption~\ref{assump:diagdom}, every column sum of
$\bM_{\mathrm{norm}}$ is at most $\mu$. Since its total sum is
$N\mu$, all column sums equal $\mu$, and therefore
\[
\bM_{\mathrm{norm}}^\top\mathbf1=\mu\mathbf1,
\qquad \mathbf S\mathbf1=\mu\mathbf1.
\]
It follows that
$\bQ\mathbf1=\mu^2(1-\mu)^{-1}\mathbf1$.
The bound $\lambda_{\max}(\bQ)\leq\mu^2/(1-\mu)$ from the proof of
\Cref{thm:PoA-spectral-exact} is thus attained. Consequently, we have that
\[
\mathrm{PoA}_{\min}
=\frac{4(1-\mu)}{(2-\mu)^2},
\]
which recovers the tight bound from \Cref{thm:PoA-tight}.
\end{example}}

{\vspace{1mm}
\begin{example}[Star network]
\label{ex:star}
Consider a star network with $N$ products in which
product~$0$ (the center) interacts with all other
products, while the remaining products do not interact
among themselves.
Formally, let
$b_{0j}=b_{j0}=\rho>0$ for $j=1,\ldots,N-1$, and
$b_{ij}=0$ for $i,j\ge1$ with $i\ne j$.
Assume {$b_{ii}=-1$} for all $i$.
Diagonal dominance requires $(N-1)\rho<1$.

The local interaction parameters are
\[
\mu_0=(N-1)\rho,
\qquad
\mu_j=\rho
\ \text{for } j\ge1,
\]
so $\max_i\mu_i=(N-1)\rho$.
The normalized interaction matrix has
$(\bM_{\mathrm{norm}})_{0j}=(\bM_{\mathrm{norm}})_{j0}
=\rho$ for $j\ge1$, and zeros elsewhere.
Its characteristic polynomial is
$\lambda^{N-2}(\lambda^2-(N-1)\rho^2)$, so its eigenvalues
are $\pm\rho\sqrt{N-1}$ and $0$ (with multiplicity
$N-2$).
Hence we have that
\[
\mu_{\mathrm{spectral}}
=
\rho\sqrt{N-1}
=
\frac{\max_i\mu_i}{\sqrt{N-1}} .
\]
For this specified star matrix, $\mathbf S=\bM_{\mathrm{norm}}$,
so the eigenvalues of $\bQ$ are $\lambda^2/(1-\lambda)$ for the
eigenvalues $\lambda$ just listed. Its largest eigenvalue is
$\rho^2(N-1)/(1-\rho\sqrt{N-1})$.
The exact spectral formula~\eqref{eq:PoA-exact} then yields
\[
\mathrm{PoA}_{\min}
=
\frac{4\bigl(1-\rho\sqrt{N-1}\bigr)}
{\bigl(2-\rho\sqrt{N-1}\bigr)^2},
\]
which, when $N>2$ and $\rho>0$, is strictly larger than the row-sum bound
\[
\frac{4\bigl(1-(N-1)\rho\bigr)}
{\bigl(2-(N-1)\rho\bigr)^2}
\]
based on $\max_i\mu_i$. For example, with $N=5$ and $\rho=0.15$, we have
$\max_i\mu_i=0.6$ and
$\mu_{\mathrm{spectral}}=0.3$.
The row-sum bound gives
$4(0.4)/(1.4)^2\approx0.816$, whereas the spectral
formula yields
$4(0.7)/(1.7)^2\approx0.969$.
Thus the spectral parameter reveals that efficiency
loss can be much smaller than suggested by the scalar
diagonal-dominance bound.
\end{example}}

{\section{Extension to Nonlinear Demand}
\label{sec:nonlinear}

The preceding analysis uses affine demand to obtain exact revenue
identities and closed-form prices. We now show that the main
price-of-anarchy guarantee continues to hold for nonlinear demand.
Own-price and cross-price sensitivities may depend on the entire price
vector, and cross-price effects need not be symmetric. The key step is
to replace the affine revenue identity by an upper bound obtained from
tangent demand.

\subsection{Demand Conditions and Equilibrium}
\label{subsec:nl-model}

For this extension, both pricing schemes use the common price set
\[
\mathcal P=\prod_{i=1}^N[0,\overline p_i],
\qquad 0<\overline p_i<\infty.
\]
Thus, unlike the unconstrained affine formulation, prices are
nonnegative and bounded. Each player still maximizes
$u_i(\bp)=p_iF_i(\bp)$, while the centralized benchmark is
$R^*=\max_{\bp\in\mathcal P}R(\bp)$, where
$R(\bp)=\sum_i p_iF_i(\bp)$. It is standard to assume that $R^*>0$.

Write the price-dependent Jacobian and its symmetric part as
\begin{equation*}
\bB(\bp)=\bigl(b_{ij}(\bp)\bigr)_{i,j=1}^N,
\qquad b_{ij}(\bp):=\frac{\partial F_i(\bp)}{\partial p_j},
\qquad
\bB_{\mathrm{s}}(\bp):=\frac{\bB(\bp)+\bB(\bp)^\top}{2}.
\end{equation*}
Here $b_{ij}(\bp)$ generalizes the constant coefficient $b_{ij}$ in
the affine model.

\vspace{1mm}
\begin{assumption}[Smooth and jointly concave demand]
\label{assump:nl-concavity}
Each $F_i$ is twice continuously differentiable on a neighborhood of
$\mathcal P$ and, for every $\bp\in\mathcal P$, it holds that
\[
F_i(\bp)\geq0,\qquad b_{ii}(\bp)<0,\qquad
\nabla^2F_i(\bp)\preceq\mathbf0.
\]
\end{assumption}

\vspace{1mm}
Joint concavity is with respect to the entire price vector, not only
a product's own price. It ensures that each demand function lies
below its tangent on the common feasible set.

\vspace{1mm}
\begin{assumption}[Two-sided dominance of the demand Jacobian]
\label{assump:nl-diagdom}
There exists $\mu\in[0,1)$ such that, for every $i$ and every
$\bp\in\mathcal P$, we have
\begin{equation}
\max\left\{
\sum_{j\ne i}|b_{ij}(\bp)|,
\sum_{j\ne i}|b_{ji}(\bp)|
\right\}
\leq\mu\bigl(-b_{ii}(\bp)\bigr).
\label{eq:nl-diagdom}
\end{equation}
\end{assumption}

\vspace{1mm}
This is the pointwise counterpart of Assumption~\ref{assump:diagdom}.
It accommodates directed substitution and complementarity effects,
including effects whose signs vary across the price domain. Define
\begin{equation*}
\begin{aligned}
\bD(\bp):=\mathrm{diag}\bigl(-b_{11}(\bp),\ldots,-b_{NN}(\bp)\bigr),\quad
\bH(\bp):=-\bB_{\mathrm{s}}(\bp),
\qquad \mathbf C(\bp):=\bB(\bp)+\bD(\bp).
\end{aligned}
\end{equation*}
The matrix $\mathbf C(\bp)$ has zero diagonal. The marginal-revenue
vector is
\[
g_i(\bp):=F_i(\bp)+p_i b_{ii}(\bp),
\qquad \mathbf g(\bp)=\bm F(\bp)-\bD(\bp)\bp.
\]
We then have the following result.
\vspace{1mm}
\begin{proposition}[Centralized optimum and nonlinear equilibrium]
\label{prop:nl-equilibrium}
Under Assumptions~\ref{assump:nl-concavity} and~\ref{assump:nl-diagdom},
the centralized revenue function is strictly concave on $\mathcal P$
and has a unique maximizer. The decentralized game admits a
pure-strategy Nash equilibrium. A vector $\bx\in\mathcal P$ is a Nash
equilibrium if and only if
\begin{equation}
\mathbf g(\bx)^\top(\bp-\bx)\leq0
\qquad\text{for every }\bp\in\mathcal P.
\label{eq:nl-equilibrium-vi}
\end{equation}
Every such equilibrium satisfies
\begin{equation}
R(\bx)\geq\bx^\top\bD(\bx)\bx.
\label{eq:nl-revenue-energy}
\end{equation}
At an interior equilibrium, $\mathbf g(\bx)=\mathbf0$ and
\eqref{eq:nl-revenue-energy} is an equality.
\end{proposition}

\vspace{1mm}
Let $\mathcal E(\bm F,\mathcal P)$ be the set of pure-strategy
Nash equilibria. We define
\[
\mathrm{PoA}(\bm F,\mathcal P)
:=\inf_{\bx\in\mathcal E(\bm F,\mathcal P)}\frac{R(\bx)}{R^*}.
\]

\subsection{The Nonlinear Price-of-Anarchy Guarantee}
\label{subsec:nl-poa}

The following comparison is the nonlinear counterpart of
\Cref{lem:revenue-gap}.

\vspace{1mm}
\begin{lemma}[Tangent-demand revenue comparison]
\label{lem:nl-tangent}
Under Assumptions~\ref{assump:nl-concavity} and~\ref{assump:nl-diagdom},
let $\bx$ be a Nash equilibrium. For every $\bp\in\mathcal P$, with
$\boldsymbol\Delta=\bp-\bx$, it holds that
\begin{equation}
R(\bp)-R(\bx)
\leq
\bigl(\mathbf C(\bx)^\top\bx\bigr)^\top\boldsymbol\Delta
-\boldsymbol\Delta^\top\bH(\bx)\boldsymbol\Delta.
\label{eq:nl-tangent-bound}
\end{equation}
\end{lemma}

\vspace{1mm}
Joint concavity gives
$F_i(\bp)\leq F_i(\bx)+\nabla F_i(\bx)^\top(\bp-\bx)$.
For a price vector $\bp$, we define the pointwise counterparts of the
spectral objects in \Cref{sec:spectral}:
\begin{equation*}
\begin{aligned}
&\bM_{\mathrm{norm}}(\bp)
:=\bD(\bp)^{-1/2}\mathbf C(\bp)\bD(\bp)^{-1/2},
&&\mathbf S(\bp)
:=\frac{\bM_{\mathrm{norm}}(\bp)+\bM_{\mathrm{norm}}(\bp)^\top}{2},\\
&\bQ(\bp)
:=\bM_{\mathrm{norm}}(\bp)
   (\bI-\mathbf S(\bp))^{-1}\bM_{\mathrm{norm}}(\bp)^\top,
&&\mu_{\mathrm{spectral}}(\bp):=\|\bM_{\mathrm{norm}}(\bp)\|_2.
\end{aligned}
\end{equation*}
Assumption~\ref{assump:nl-diagdom} implies
$\mu_{\mathrm{spectral}}(\bp)\leq\mu<1$ and
$\bI-\mathbf S(\bp)\succ0$. Then, we have the following result.

\vspace{1mm}
\begin{theorem}[Price of anarchy under nonlinear demand]
\label{thm:nl-poa}
Under Assumptions~\ref{assump:nl-concavity} and~\ref{assump:nl-diagdom},
every pure-strategy Nash equilibrium $\bx$ has $R(\bx)>0$ and satisfies
\begin{equation}
R^*-R(\bx)
\leq\frac14\bx^\top\mathbf C(\bx)\bH(\bx)^{-1}
                         \mathbf C(\bx)^\top\bx.
\label{eq:nl-gap-certificate}
\end{equation}
Consequently, we have that
\begin{equation}
\begin{aligned}
\frac{R(\bx)}{R^*}
&\geq
\left(1+\frac{\bx^\top\mathbf C(\bx)\bH(\bx)^{-1}
\mathbf C(\bx)^\top\bx}{4R(\bx)}\right)^{-1}\geq\frac{4}{4+\lambda_{\max}(\bQ(\bx))}\\
&\geq\frac{4(1-\mu_{\mathrm{spectral}}(\bx))}
              {(2-\mu_{\mathrm{spectral}}(\bx))^2}
\geq\frac{4(1-\mu)}{(2-\mu)^2}.
\end{aligned}
\label{eq:nl-poa-chain}
\end{equation}
In particular, it holds that
$\mathrm{PoA}(\bm F,\mathcal P)\geq4(1-\mu)/(2-\mu)^2$.
\end{theorem}
Note that \Cref{thm:nl-poa} extends the ratio in \Cref{thm:PoA-mu-bound} from the linear case to the nonlinear case. The bound coincides with the affine guarantee. 

\subsection{Uniqueness and Nonlinear Examples}
\label{subsec:nl-uniqueness}

Demand dominance controls the revenue loss, whereas a separate
condition on marginal revenues provides uniqueness. Write
$\partial_i b_{ij}(\bp)=\partial b_{ij}(\bp)/\partial p_i$.
Then, we have that
\[
\partial_jg_i(\bp)=b_{ij}(\bp)+p_i\partial_i b_{ij}(\bp)
\quad(j\ne i),
\qquad
\partial_i g_i(\bp)=2b_{ii}(\bp)+p_i\partial_i b_{ii}(\bp)<0.
\]
The equality of mixed partial derivatives used here follows from
smoothness of $F_i$.

\vspace{1mm}
\begin{proposition}[Uniqueness under marginal-revenue dominance]
\label{prop:nl-uniqueness}
Suppose Assumptions~\ref{assump:nl-concavity} and~\ref{assump:nl-diagdom}
hold. If there exists $\kappa\in[0,1)$ such that, for every $i$ and
$\bp\in\mathcal P$, it holds
\begin{equation}
\sum_{j\ne i}|b_{ij}(\bp)+p_i\partial_i b_{ij}(\bp)|
\leq\kappa\bigl[-2b_{ii}(\bp)-p_i\partial_i b_{ii}(\bp)\bigr],
\label{eq:nl-marginal-dominance}
\end{equation}
then the Nash equilibrium is
unique, and simultaneous best-response iterates converge to it from every initial vector in $\mathcal P$.
\end{proposition}

\vspace{1mm}
\begin{example}[Nonlinear own-price effects with directed interactions]
\label{ex:nl-own}
Consider
\begin{equation*}
F_i(\bp)=a_i+\sum_{j=1}^N b_{ij}p_j-\gamma_i p_i^2,
\qquad \gamma_i\geq0,
\end{equation*}
where the constant matrix $\bB=(b_{ij})$ satisfies the affine
own-price and two-sided dominance assumptions. Then $b_{ii}(\bp)=b_{ii}-2\gamma_i p_i<0$, while
$b_{ij}(\bp)=b_{ij}$ for $j\ne i$, so the Jacobian dominance is preserved.
Demand is jointly concave, and it is nonnegative
provided that
\[
a_i\geq(-b_{ii})\overline p_i+\gamma_i\overline p_i^2
       +\sum_{\substack{j\ne i:\ b_{ij}<0}}|b_{ij}|\overline p_j.
\]
Moreover, \eqref{eq:nl-marginal-dominance} holds with
$\kappa=\mu/2$, because its left side is $\sum_{j\ne i}|b_{ij}|$
and the bracket on its right side is
$-2b_{ii}+6\gamma_i p_i\geq2(-b_{ii})$.
Thus this family has a unique equilibrium and satisfies the same
price-of-anarchy guarantee for any nonnegative curvature parameters
consistent with demand nonnegativity.
\end{example}

\vspace{1mm}
\begin{example}[Nonlinear and asymmetric cross-price effects]
\label{ex:nl-cross}
Let $\bw\in\R_+^N$ and $\gamma_i\geq0$, and consider
\begin{equation*}
F_i(\bp)=a_i+\sum_{j=1}^N b_{ij}p_j
                   -\gamma_i(\bw^\top\bp)^2,
\qquad b_{ii}<0.
\end{equation*}
Here, we have that
\[
b_{ij}(\bp)=b_{ij}-2\gamma_iw_j(\bw^\top\bp),
\qquad
\nabla^2F_i(\bp)=-2\gamma_i\bw\bw^\top\preceq\mathbf0.
\]
\Cref{thm:nl-poa} applies whenever demand is nonnegative and
\eqref{eq:nl-diagdom} holds.
\end{example}

\subsection{A Curvature-Adjusted Extension}
\label{subsec:nl-curvature}

Joint demand concavity can be weakened by bounding the positive part
of the tangent remainder. Retain smoothness, demand nonnegativity,
negative own-price effects, and Assumption~\ref{assump:nl-diagdom},
but replace joint demand concavity by a weaker condition that
\begin{equation}
2b_{ii}(\bp)+p_i\partial_i b_{ii}(\bp)<0
\qquad\text{for every }i,\bp\in\mathcal P.
\label{eq:nl-own-revenue-concavity}
\end{equation}
This ensures strict concavity of each player's revenue in its own
price and hence equilibrium existence, but need not imply concavity
of centralized revenue. For an equilibrium $\bx$, define
\[
E_{\bx}(\bp):=\sum_i p_i
\bigl[F_i(\bp)-F_i(\bx)
              -\nabla F_i(\bx)^\top(\bp-\bx)\bigr].
\]
We have the following result.
\vspace{1mm}
\begin{proposition}[Efficiency with bounded positive curvature]
\label{prop:nl-curvature}
Under the condition \eqref{eq:nl-own-revenue-concavity}, let $\bx$ be a Nash
equilibrium. If there exists $\eta\in[0,1-\mu)$ such that
\begin{equation}
E_{\bx}(\bp)\leq\eta(\bp-\bx)^\top\bD(\bx)(\bp-\bx)
\qquad\text{for every }\bp\in\mathcal P,
\label{eq:nl-remainder-control}
\end{equation}
then, it holds that
\begin{equation}
\frac{R(\bx)}{R^*}
\geq\frac{4(1-\mu-\eta)}{4(1-\mu-\eta)+\mu^2}.
\label{eq:nl-curvature-poa}
\end{equation}
If the same $\eta$ works at every equilibrium, the bound also holds
for $\mathrm{PoA}(\bm F,\mathcal P)$.
\end{proposition}

\vspace{1mm}
A sufficient condition is obtained by defining
\[
\underline{\bD}:=\mathrm{diag}\left(
\min_{\bp\in\mathcal P}[-b_{11}(\bp)],\ldots,
\min_{\bp\in\mathcal P}[-b_{NN}(\bp)]\right)\succ0
\]
and finding constants $L_i\geq0$ for which the following inequalities hold
\begin{equation*}
\nabla^2F_i(\bp)\preceq L_i\underline{\bD}
\quad\text{for every }i,\bp\in\mathcal P,
\qquad
\eta:=\frac12\sum_i\overline p_iL_i<1-\mu.
\end{equation*}
Then we obtain \eqref{eq:nl-remainder-control} at every
equilibrium. Note that jointly concave demand corresponds to $L_i=0$ and
$\eta=0$, which recovers \Cref{thm:nl-poa}.}

\section{Concluding Remarks}

In this paper, we study the revenue lost when the prices of
interrelated products are set by decision makers who each maximize the
revenue of a single product. We model this as a noncooperative game
under linear demand in which cross-price effects may be substitutes or
complements and need not be symmetric. Under two-sided diagonal
dominance, the game has a unique Nash equilibrium, and equilibrium
revenue is at least a fraction $4(1-\mu)/(2-\mu)^2$ of the centralized
optimum, where $\mu$ measures the aggregate strength of cross-price
effects relative to own-price effects. This bound was known for
symmetric gross substitutes; we show that it remains valid for signed
and asymmetric interactions, and that it is tight. Decentralized
pricing is therefore nearly efficient when products are weakly coupled,
while the worst-case loss grows rapidly as $\mu$ approaches one.

Because asymmetry rules out the simultaneous diagonalization used in
existing analyses, our argument instead evaluates the revenue gap at
the equilibrium itself. This yields an exact worst-case formula for
each demand matrix in terms of a normalized interaction matrix. The
formula shows that the pattern of interactions, not only their
aggregate strength, determines the loss: at the same maximum local
interaction parameter, a star network loses much less revenue in the
worst case than uniformly connected products. Because the argument does
not rely on closed-form equilibrium prices, it also extends to jointly
concave nonlinear demand, where it applies to every pure-strategy
equilibrium, and yields a curvature-adjusted bound when demand is not
jointly concave.

These results indicate when coordination is worth its cost. Suppose
that, for every product, both the total absolute effect of other prices
on its demand and the total absolute effect of its price on the demand
for other products are at most half of its own-price effect. Then
product-level pricing retains at least $88.9\%$ of centralized revenue,
so the gain from building a joint demand model and coordinating prices
is limited. This condition involves only aggregate row and column sums
of cross-price effects, so it can be assessed without estimating the
full matrix of cross-price effects, at least when products are
substitutes. When products are strongly coupled, the exact formula
shows how the loss depends on the pattern of interactions, which can
guide where coordination is most valuable. On platforms, the same
comparison bounds the revenue that coordination devices, such as
pricing tools or price recommendations, could recover.

Future research may incorporate inventory or capacity constraints,
extend the analysis to firms that own several products under asymmetric
demand, analyze finite-time learning dynamics within this game, or
study the outcomes of learning algorithms that do not converge to the
Nash equilibrium, such as those that fit misspecified demand models
\citep{cooper2015learning, hansen2021algorithmic}.

\bibliographystyle{abbrvnat}
\bibliography{bibliography}

\clearpage

\OneAndAHalfSpacedXI

%
%
%

\begin{APPENDICES}
\crefalias{section}{appendix}
\section{Missing Proofs in \Cref{sec:model}}
\subsection{Proof of \Cref{lem:B-negdef}}
Write $\bH=-\bB_{\mathrm{s}}$. Its diagonal entries are $-b_{ii}>0$, and
\[
\sum_{j\ne i}|H_{ij}|
\leq\frac12\sum_{j\ne i}(|b_{ij}|+|b_{ji}|)\leq\mu(-b_{ii}).
\]
For every real vector $\bv$, symmetry and $2|v_iv_j|\leq v_i^2+v_j^2$
give
\[
{\left|\bv^\top(\bH-\bD)\bv\right|
\leq\sum_i v_i^2\sum_{j\ne i}|H_{ij}|
\leq\mu\bv^\top\bD\bv.}
\]
{Thus it holds that $(1-\mu)\bD\preceq\bH\preceq(1+\mu)\bD$, proving positive
definiteness of $\bH$ and negative definiteness of $\bB_{\mathrm{s}}$.
It is direct to see that the standard Gershgorin statement is (see, e.g. Theorem 6.1.1 of \cite{horn2012matrix}) is retained.}
\begin{claim}\label{claim:eigenvalue}
{Every eigenvalue} $\lambda$ {of} $\bB$ {belongs to a complex disc such that
$|\lambda-b_{ii}|\leq\sum_{j\ne i}|b_{ij}|$ for some $i$.}
\end{claim}
{Under row dominance each such disc lies strictly in the left half-plane.
Hence every eigenvalue of} $\bB$ {has negative real part and} $\bB$ {is
invertible. This holds even without asserting that its eigenvalues are real}.

\section{Missing Proofs in \Cref{sec:DecentralizedPricing}}
\subsection{Proof of \Cref{thm:central-opt}}

Recall that
\[
    {R(\bp) = \ba^\top \bp + \bp^\top \bB \bp = \ba^\top \bp + \bp^\top \bB_{\mathrm{s}} \bp.}
\]
We compute the gradient $\nabla R(\bp) \in \R^N$, which is the vector of partial derivatives. It is direct to see that the gradient of the linear term $\ba^\top \bp$ with respect to $\bp$ is $\ba$. For the quadratic term {$\bp^\top \bB_{\mathrm{s}} \bp$}, we use the following standard result:
{Since $\bB_{\mathrm{s}}$} is symmetric, we have that {$\nabla_{\bp}(\bp^\top \bB_{\mathrm{s}} \bp) = 2 \bB_{\mathrm{s}} \bp$}. Combining these results, we have that
\begin{equation}
    {\nabla R(\bp) = \ba + 2 \bB_{\mathrm{s}} \bp.
    \label{eq:grad-R}}
\end{equation}
The Hessian matrix $\nabla^2 R(\bp) \in \R^{N \times N}$ is the matrix of second-order partial derivatives.
Differentiating \eqref{eq:grad-R} with respect to $\bp$, we obtain that
\[
    \nabla^2 R(\bp) = 2 \bB_{\mathrm{s}}.
\]
Note that the Hessian is constant (does not depend on $\bp$), and is negative definitive (following \Cref{lem:B-negdef}). Therefore, we know that the function $R(\bp)$ is a strictly concave function and enjoys a unique optimal solution. 

For a strictly concave function on $\R^N$, a point $\bp^*$ is the unique global maximizer if and only if it satisfies the first-order necessary condition $\nabla R(\bp^*) = \mathbf{0}$.
This is because strict concavity ensures that any critical point is a global maximum, and there can be at most one such point.
Setting $\nabla R(\bp^*) = \mathbf{0}$ using \eqref{eq:grad-R}, we have that
\[
    {\ba + 2 \bB_{\mathrm{s}} \bp^* = \mathbf{0}.}
\]
Rearranging the term, we have that
\[
    {2 \bB_{\mathrm{s}} \bp^* = -\ba.}
\]
Since {$\bB_{\mathrm{s}}$} is invertible (by Lemma~\ref{lem:B-negdef}), we can multiply both sides on the left by {$\bB_{\mathrm{s}}^{-1}$} and finally get that
\[
    {\bp^* = -\frac{1}{2} \bB_{\mathrm{s}}^{-1} \ba.}
\]
This proves the first part.

We now substitute {$\bp^* = -\frac{1}{2} \bB_{\mathrm{s}}^{-1} \ba$} into the revenue function {$R(\bp) = \ba^\top \bp + \bp^\top \bB_{\mathrm{s}} \bp$}. We have that
\begin{align*}
R(\bp^*)
    &= \ba^\top \bp^* + (\bp^*)^\top \bB_{\mathrm{s}} \bp^*
         = -\frac{1}{2} \ba^\top \bB_{\mathrm{s}}^{-1} \ba + \frac{1}{4} \ba^\top \bB_{\mathrm{s}}^{-1} \ba
         \\
    &{= \left( -\frac{1}{2} + \frac{1}{4} \right) \ba^\top \bB_{\mathrm{s}}^{-1} \ba
        \notag} = -\frac{1}{4} \ba^\top \bB_{\mathrm{s}}^{-1} \ba.   
\end{align*}
This proves the second part and completes the proof.


\subsection{Proof of \Cref{prop:FOC-Nash}}
\noindent$(\Rightarrow)$
Suppose $\bp^{\mathrm{NE}}$ is a Nash equilibrium.
By Definition~\ref{def:Nash-equilibrium}, for each player $i$, the price $p_i^{\mathrm{NE}}$ maximizes $u_i(p_i; \bp^{\mathrm{NE}}_{-i})$ over all $p_i \in \R$.
The unique maximizer satisfies the first-order condition \eqref{eq:FOC-player-i}, which, by \eqref{eq:best-response-explicit}, is
\[
    a_i + 2 b_{ii} p_i^{\mathrm{NE}} + \sum_{j \neq i} b_{ij} p_j^{\mathrm{NE}} = 0.
\]

\noindent$(\Leftarrow)$
Conversely, suppose $\bp^{\mathrm{NE}}$ satisfies \eqref{eq:FOC-NE-coord} for all $i$.
For each $i$, this means $p_i^{\mathrm{NE}}$ satisfies the first-order condition for the problem $\max_{p_i} u_i(p_i; \bp^{\mathrm{NE}}_{-i})$.
Since $u_i(\cdot; \bp^{\mathrm{NE}}_{-i})$ is strictly concave, any critical point is the unique global maximizer.
Therefore, $p_i^{\mathrm{NE}}$ is the best response to $\bp^{\mathrm{NE}}_{-i}$.
Since this holds for all $i$, $\bp^{\mathrm{NE}}$ is a Nash equilibrium.

{\subsection{Proof of \Cref{lem:A-negdef}}

By definition, it holds
\[
\bA^{\mathrm{NE}}=\bB-\bD,
\qquad
\frac{\bA^{\mathrm{NE}}+(\bA^{\mathrm{NE}})^\top}{2}
=\bB_{\mathrm{s}}-\bD.
\]
We first verify that this symmetric part is negative definite.
For every $\bv\in\R^N$, two-sided diagonal dominance gives
\begin{align*}
\bv^\top(\bB_{\mathrm{s}}-\bD)\bv
&=2\sum_i b_{ii}v_i^2
  +\sum_{i<j}(b_{ij}+b_{ji})v_iv_j\\
&\leq -2\sum_i(-b_{ii})v_i^2
  +\frac12\sum_{i<j}(|b_{ij}|+|b_{ji}|)(v_i^2+v_j^2)\\
&= -2\sum_i(-b_{ii})v_i^2
  +\frac12\sum_i v_i^2
      \left(\sum_{j\ne i}|b_{ij}|+\sum_{j\ne i}|b_{ji}|\right)\\
&\leq -(2-\mu)\sum_i(-b_{ii})v_i^2.
\end{align*}
The first inequality uses $2|v_iv_j|\leq v_i^2+v_j^2$.
Since $-b_{ii}>0$ and $2-\mu>0$, the last expression is strictly
negative whenever $\bv\ne\mathbf0$.

To deduce invertibility, suppose that
$\bA^{\mathrm{NE}}\bv=\mathbf0$ for a real vector $\bv$.
Then
\[
0=\bv^\top\bA^{\mathrm{NE}}\bv
 =\bv^\top\frac{\bA^{\mathrm{NE}}+(\bA^{\mathrm{NE}})^\top}{2}\bv.
\]
Negative definiteness forces $\bv=\mathbf0$. Thus the real square
matrix $\bA^{\mathrm{NE}}$ has a trivial kernel and is invertible.
This proves both conclusions without imposing symmetry on
$\bA^{\mathrm{NE}}$.}

\subsection{Proof of \Cref{thm:NE-characterization}}
The proof consists of three parts: (i) any Nash equilibrium must satisfy a certain linear system, (ii) this linear system has a unique solution, and (iii) this solution is indeed a Nash equilibrium.

\medskip
\noindent\textbf{Part (i): Any Nash equilibrium satisfies the linear system.}
Let $\bp^{\mathrm{NE}}$ be a Nash equilibrium.
By Proposition~\ref{prop:FOC-Nash}, $\bp^{\mathrm{NE}}$ satisfies the first-order conditions \eqref{eq:FOC-NE-coord} for all $i$, which is equivalent to the linear system
\begin{equation}
    \bA^{\mathrm{NE}} \bp^{\mathrm{NE}} = -\ba.
    \tag{\ref{eq:NE-system}}
\end{equation}

\medskip
\noindent\textbf{Part (ii): The linear system has a unique solution.}
By \Cref{lem:A-negdef}, $\bA^{\mathrm{NE}}$ is invertible.
Therefore, the linear system $\bA^{\mathrm{NE}} \bp = -\ba$ has a unique solution:
\[
    \bp = -(\bA^{\mathrm{NE}})^{-1} \ba.
\]

Combining Parts (i) and (ii): if a Nash equilibrium exists, it must be unique and equal to $-(\bA^{\mathrm{NE}})^{-1} \ba$.

\medskip
\noindent\textbf{Part (iii): The solution of the linear system is a Nash equilibrium.}
Let $\tilde{\bp} = -(\bA^{\mathrm{NE}})^{-1} \ba$ be the unique solution of the linear system \eqref{eq:NE-system}.
We must show that $\tilde{\bp}$ is a Nash equilibrium, i.e., for each player $i$, $\tilde{p}_i$ is a best response to $\tilde{\bp}_{-i}$.

Fix any player $i \in \{1, \ldots, N\}$.
Consider player $i$'s optimization problem given $\tilde{\bp}_{-i}$:
\[
    \max_{p_i \in \R} u_i(p_i; \tilde{\bp}_{-i}).
\]
From the {payoff expression \eqref{eq:ui-expanded}}, we can see that the function $u_i(\cdot; \tilde{\bp}_{-i})$ is strictly concave in $p_i$, and the unique maximizer is characterized by the first-order condition $\frac{\partial u_i}{\partial p_i}(p_i; \tilde{\bp}_{-i}) = 0$.
The first-order condition is exactly
\[
    a_i + 2 b_{ii} p_i + \sum_{j \neq i} b_{ij} \tilde{p}_j = 0,
\]
which is the system described in \eqref{eq:NE-system}. From the {invertibility} of the matrix $\bA^{\mathrm{NE}}$ established in \Cref{lem:A-negdef}, we know that $p_i = \tilde{p}_i$, satisfying the first-order condition for player $i$'s problem when opponents play $\tilde{\bp}_{-i}$.
By strict concavity, $\tilde{p}_i$ is the unique global maximizer, hence the best response to $\tilde{\bp}_{-i}$.
Since this holds for every player $i$, $\tilde{\bp}$ is a Nash equilibrium.

As a result, we have shown that:
\begin{itemize}
    \item Any Nash equilibrium must equal $-(\bA^{\mathrm{NE}})^{-1} \ba$ (by Parts (i) and (ii)).
    \item The vector $-(\bA^{\mathrm{NE}})^{-1} \ba$ is indeed a Nash equilibrium (by Part (iii)).
\end{itemize}
Therefore, the decentralized pricing game has a unique Nash equilibrium, given by $\bp^{\mathrm{NE}} = -(\bA^{\mathrm{NE}})^{-1} \ba$, which completes our proof.

\subsection{Proof of \Cref{prop:PoA-ratio}}
From Theorem~\ref{thm:central-opt}, {$R(\bp^*) = -\frac{1}{4} \ba^\top \bB_{\mathrm{s}}^{-1} \ba$}.
We now establish that $R(\bp^{\mathrm{NE}}) = -\frac{1}{4} \ba^\top \bK \ba$.
{By \Cref{def:K-matrix},} the matrix $\bK$ is symmetric negative definite{,
since $\bD\succ0$ and $\bA^{\mathrm{NE}}$ is invertible}.
We now derive the expression for $R(\bp^{\mathrm{NE}})$.
For any price vector $\bp$, it holds that
\[
    {R(\bp) = \ba^\top \bp + \bp^\top \bB_{\mathrm{s}} \bp.}
\]
We substitute $\bp = \bp^{\mathrm{NE}} = -(\bA^{\mathrm{NE}})^{-1} \ba$ and compute each term. For the first term, we have
\begin{align*}
    \ba^\top \bp^{\mathrm{NE}}
    = \ba^\top \left( -(\bA^{\mathrm{NE}})^{-1} \ba \right)
        = -\ba^\top (\bA^{\mathrm{NE}})^{-1} \ba.
\end{align*}
For the second term, we have that
\begin{align*}
    {(\bp^{\mathrm{NE}})^\top \bB_{\mathrm{s}} \bp^{\mathrm{NE}}}
    &{= \left( -(\bA^{\mathrm{NE}})^{-1} \ba \right)^\top \bB_{\mathrm{s}} \left( -(\bA^{\mathrm{NE}})^{-1} \ba \right)
        \notag}{= \ba^\top \left( (\bA^{\mathrm{NE}})^{-1} \right)^\top \bB_{\mathrm{s}} (\bA^{\mathrm{NE}})^{-1} \ba
         \notag}\\
    &= \ba^\top (\bA^{\mathrm{NE}})^{-\top} \bB_{\mathrm{s}} (\bA^{\mathrm{NE}})^{-1} \ba.
\end{align*}
Combining both terms {and using
$\bD=\bB_{\mathrm{s}}-\tfrac12(\bA^{\mathrm{NE}}+(\bA^{\mathrm{NE}})^\top)$,
we obtain}
\begin{align*}
{R(\bp^{\mathrm{NE}})}
&{=-\ba^\top(\bA^{\mathrm{NE}})^{-1}\ba
 +\ba^\top(\bA^{\mathrm{NE}})^{-\top}\bB_{\mathrm{s}}
             (\bA^{\mathrm{NE}})^{-1}\ba\notag}\\
&{=\ba^\top(\bA^{\mathrm{NE}})^{-\top}
 \left(\bB_{\mathrm{s}}-\frac{\bA^{\mathrm{NE}}+
                                  (\bA^{\mathrm{NE}})^\top}{2}\right)
 (\bA^{\mathrm{NE}})^{-1}\ba\notag}\\
&=\ba^\top(\bA^{\mathrm{NE}})^{-\top}\bD
             (\bA^{\mathrm{NE}})^{-1}\ba
 =-\frac14\ba^\top\bK\ba.
\end{align*}
{Here we used that a scalar quadratic form is unchanged by replacing
its matrix by its symmetric part. Combining this identity with
$R(\bp^*)=-\tfrac14\ba^\top\bB_{\mathrm{s}}^{-1}\ba$ proves
\eqref{eq:PoA-ratio}.}

{\subsection{Proof of \Cref{lem:M-matrix}}

Recall that
\[
\bH=-\bB_{\mathrm{s}},\qquad
\bG=\bD-\bB=-\bA^{\mathrm{NE}},\qquad
\bY=\bH^{1/2}\bG^{-1}\bH^{1/2}.
\]
The assumptions imply $\bD\succ0$ and $\bH\succ0$, and $\bG$ is
invertible. Since $\tilde{\bL}=\bH^{-1}$ and
$\tilde{\bK}=4\bG^{-\top}\bD\bG^{-1}$, the definition of $\bM$ gives
\[
\bM=4\bH^{1/2}\bG^{-\top}\bD\bG^{-1}\bH^{1/2}.
\]
This matrix is symmetric. For every nonzero $\bv$,
\[
\bv^\top\bM\bv
=4\bigl(\bG^{-1}\bH^{1/2}\bv\bigr)^\top
     \bD\bigl(\bG^{-1}\bH^{1/2}\bv\bigr)>0,
\]
because $\bG^{-1}\bH^{1/2}\bv\ne\mathbf0$. Hence $\bM\succ0$.

Next, the definitions of $\bG$ and $\bH$ imply
\[
\frac{\bG+\bG^\top}{2}=\bD+\bH,
\qquad
\bD=\frac{\bG+\bG^\top}{2}-\bH.
\]
Multiplying the latter identity on the left by $\bG^{-\top}$ and
on the right by $\bG^{-1}$ yields
\begin{align*}
\bG^{-\top}\bD\bG^{-1}
&=\frac12\bG^{-\top}\bG\bG^{-1}
  +\frac12\bG^{-\top}\bG^\top\bG^{-1}
  -\bG^{-\top}\bH\bG^{-1}\\
&=\frac12(\bG^{-\top}+\bG^{-1})
  -\bG^{-\top}\bH\bG^{-1}.
\end{align*}
Consequently,
\begin{align*}
\bM
&=2\bH^{1/2}(\bG^{-\top}+\bG^{-1})\bH^{1/2}
  -4\bH^{1/2}\bG^{-\top}\bH\bG^{-1}\bH^{1/2}\\
&=2(\bY+\bY^\top)-4\bY^\top\bY.
\end{align*}
Here
$\bY^\top\bY=\bH^{1/2}\bG^{-\top}\bH\bG^{-1}\bH^{1/2}$;
no interchange of noncommuting factors is used.
Finally, direct multiplication gives
\[
(\bI-2\bY)^\top(\bI-2\bY)
=\bI-2(\bY+\bY^\top)+4\bY^\top\bY
=\bI-\bM.
\]
The left-hand side is positive semidefinite, so
$\mathbf0\prec\bM\preceq\bI$.
If $\bM\bv=\lambda\bv$ with $\bv\ne\mathbf0$, these inequalities
imply $0<\lambda\|\bv\|_2^2\leq\|\bv\|_2^2$.
Thus every eigenvalue of $\bM$ belongs to $(0,1]$.}

{\subsection{Proof of \Cref{thm:PoA-spectral}}

The revenue formulas and the definitions of $\tilde{\bK}$ and
$\tilde{\bL}$ give, for every $\ba\ne\mathbf0$,
\[
R(\bp^{\mathrm{NE}})=\frac14\ba^\top\tilde{\bK}\ba,
\qquad
R(\bp^*)=\frac14\ba^\top\tilde{\bL}\ba,
\qquad
\mathrm{PoA}(\ba)=
\frac{\ba^\top\tilde{\bK}\ba}{\ba^\top\tilde{\bL}\ba}.
\]
Because $\tilde{\bL}\succ0$, its symmetric square root is invertible.
Set $\mathbf x=\tilde{\bL}^{1/2}\ba$, so that
$\ba=\tilde{\bL}^{-1/2}\mathbf x$ and
$\mathbf x\ne\mathbf0$. Substitution gives
\begin{align*}
\ba^\top\tilde{\bK}\ba
&=\mathbf x^\top\tilde{\bL}^{-1/2}
               \tilde{\bK}\tilde{\bL}^{-1/2}\mathbf x
 =\mathbf x^\top\bM\mathbf x,\\
\ba^\top\tilde{\bL}\ba
&=\mathbf x^\top\tilde{\bL}^{-1/2}
               \tilde{\bL}\tilde{\bL}^{-1/2}\mathbf x
 =\mathbf x^\top\mathbf x.
\end{align*}
This proves the claimed Rayleigh-quotient representation.

To compute its extrema, choose an orthonormal eigenbasis
$\bv_1,\ldots,\bv_N$ of the symmetric matrix $\bM$, with eigenvalues
$\lambda_1\leq\cdots\leq\lambda_N$. Write
$\mathbf x=\sum_{i=1}^N\alpha_i\bv_i$.
Orthogonality implies
\[
\frac{\mathbf x^\top\bM\mathbf x}{\mathbf x^\top\mathbf x}
=\frac{\sum_i\lambda_i\alpha_i^2}{\sum_i\alpha_i^2}.
\]
The denominator is positive, and this expression is a weighted
average of the eigenvalues with nonnegative weights.
It lies between $\lambda_1$ and $\lambda_N$; taking
$\mathbf x=\bv_1$ or $\mathbf x=\bv_N$ attains the respective endpoint.
Since $\ba\mapsto\mathbf x$ is a bijection on
$\R^N\setminus\{\mathbf0\}$, the corresponding intercepts
$\ba=\tilde{\bL}^{-1/2}\bv_1$ and
$\ba=\tilde{\bL}^{-1/2}\bv_N$ are admissible and nonzero.
Therefore,
\[
\inf_{\ba\ne\mathbf0}\mathrm{PoA}(\ba)=\lambda_{\min}(\bM),
\qquad
\sup_{\ba\ne\mathbf0}\mathrm{PoA}(\ba)=\lambda_{\max}(\bM),
\]
and both extrema are attained in the unrestricted intercept domain.}

{\subsection{Proof of \Cref{lem:interaction-norm}}

Let $\mathbf C=\bB+\bD$ and
$\mathbf T=\bD^{-1/2}\mathbf C\bD^{-1/2}$.
Then $C_{ii}=0$, and the assumptions give, for every $i$,
\[
\sum_j|C_{ij}|\leq\mu(-b_{ii}),
\qquad
\sum_j|C_{ji}|\leq\mu(-b_{ii}).
\]
We first establish the norm estimate, keeping the row and column
uses of diagonal dominance explicit. For any $\bv\in\R^N$,
weighted Cauchy--Schwarz gives
\[
|(\mathbf C\bv)_i|^2
\leq\left(\sum_j|C_{ij}||v_j|\right)^2
\leq\left(\sum_j|C_{ij}|\right)
     \left(\sum_j|C_{ij}|v_j^2\right).
\]
This remains valid when a row is zero. Dividing by $-b_{ii}>0$
and summing, we obtain
\begin{align*}
\|\bD^{-1/2}\mathbf C\bv\|_2^2
&=\sum_i\frac{|(\mathbf C\bv)_i|^2}{-b_{ii}}\leq\sum_i\frac{\sum_j|C_{ij}|}{-b_{ii}}
                    \sum_j|C_{ij}|v_j^2\leq\mu\sum_jv_j^2\sum_i|C_{ij}|\\
&\leq\mu^2\sum_j(-b_{jj})v_j^2
 =\mu^2\|\bD^{1/2}\bv\|_2^2.
\end{align*}
The second inequality uses the row bounds; the third uses the
column bounds. Set $\bv=\bD^{-1/2}\bz$ to conclude that
$\|\mathbf T\bz\|_2^2\leq\mu^2\|\bz\|_2^2$ for every $\bz$,
and hence $\|\mathbf T\|_2\leq\mu$.

The transpose estimate follows from the following calculation,
which uses the column bounds first and the row bounds second:
\begin{align*}
\|\bD^{-1/2}\mathbf C^\top\bv\|_2^2
=\sum_i\frac{|\sum_j C_{ji}v_j|^2}{-b_{ii}}\leq\sum_i\frac{\sum_j|C_{ji}|}{-b_{ii}}
                    \sum_j|C_{ji}|v_j^2\leq\mu\sum_jv_j^2\sum_i|C_{ji}|\leq\mu^2\sum_j(-b_{jj})v_j^2.
\end{align*}
Equivalently, for every $\bv$,
\[
\bv^\top\mathbf C\bD^{-1}\mathbf C^\top\bv
\leq\mu^2\bv^\top\bD\bv,
\]
so $\mathbf C\bD^{-1}\mathbf C^\top\preceq\mu^2\bD$.

To control revenue curvature, define
$\mathbf S=(\mathbf T+\mathbf T^\top)/2$.
Then $\|\mathbf S\|_2\leq\|\mathbf T\|_2\leq\mu$ and
\[
\bH=-\bB_{\mathrm{s}}
=\bD^{1/2}(\bI-\mathbf S)\bD^{1/2}.
\]
Every eigenvalue of $\mathbf S$ lies in $[-\mu,\mu]$, so
$\bI-\mathbf S\succ0$ and
$(\bI-\mathbf S)^{-1}\preceq(1-\mu)^{-1}\bI$.
It follows that
\[
\bH^{-1}
=\bD^{-1/2}(\bI-\mathbf S)^{-1}\bD^{-1/2}
\preceq\frac1{1-\mu}\bD^{-1}.
\]
For an arbitrary $\bv$, apply this inequality to $\mathbf C^\top\bv$:
\begin{align*}
\bv^\top\mathbf C\bH^{-1}\mathbf C^\top\bv
\leq\frac1{1-\mu}
       \bv^\top\mathbf C\bD^{-1}\mathbf C^\top\bv\leq\frac{\mu^2}{1-\mu}\bv^\top\bD\bv.
\end{align*}
The desired matrix inequality follows because this holds for every
real vector $\bv$.}

{\subsection{Proof of \Cref{thm:PoA-mu-bound}}

Fix $\ba\ne\mathbf0$ and write
\[
\bx=\bp^{\mathrm{NE}},\qquad
\mathbf C=\bB+\bD,\qquad
\bH=-\bB_{\mathrm{s}},\qquad
\bG=\bD-\bB.
\]
The equilibrium equations are $\bG\bx=\ba$. Thus
$\bm F(\bx)=\ba+\bB\bx=\bD\bx$, and
\[
R(\bx)=\bx^\top\bD\bx>0.
\]
Here $\bx\ne\mathbf0$ because $\ba\ne\mathbf0$ and $\bG$ is invertible.
Also,
\[
\bG-2\bH
=\bD-\bB+(\bB+\bB^\top)
=\bD+\bB^\top=\mathbf C^\top.
\]
Hence $\ba-2\bH\bx=\mathbf C^\top\bx$.

For any $\bp\in\R^N$, put $\boldsymbol\Delta=\bp-\bx$.
Using $R(\bp)=\ba^\top\bp-\bp^\top\bH\bp$ and symmetry of $\bH$,
\begin{align*}
R(\bp)-R(\bx)
&=\ba^\top\boldsymbol\Delta
 -\bx^\top\bH\boldsymbol\Delta
 -\boldsymbol\Delta^\top\bH\bx
 -\boldsymbol\Delta^\top\bH\boldsymbol\Delta=(\mathbf C^\top\bx)^\top\boldsymbol\Delta
 -\boldsymbol\Delta^\top\bH\boldsymbol\Delta\\
&=\frac14\bx^\top\mathbf C\bH^{-1}\mathbf C^\top\bx-
 \left(\boldsymbol\Delta-\frac12\bH^{-1}\mathbf C^\top\bx\right)^\top
 \bH
 \left(\boldsymbol\Delta-\frac12\bH^{-1}\mathbf C^\top\bx\right).
\end{align*}
Since $\bH\succ0$, the last quadratic term is nonnegative and
vanishes precisely at
$\boldsymbol\Delta=\frac12\bH^{-1}\mathbf C^\top\bx$.
This displacement is feasible because the centralized domain is
$\R^N$. Moreover,
\[
\bx+\frac12\bH^{-1}\mathbf C^\top\bx
=\frac12\bH^{-1}(2\bH+\mathbf C^\top)\bx
=\frac12\bH^{-1}\ba=\bp^*.
\]
Therefore,
\[
R(\bp^*)-R(\bx)
=\frac14\bx^\top\mathbf C\bH^{-1}\mathbf C^\top\bx.
\]

The assumptions of \Cref{lem:interaction-norm} hold for this same
$\bB$, and that lemma states the matrix inequality
\[
\mathbf C\bH^{-1}\mathbf C^\top
\preceq\frac{\mu^2}{1-\mu}\bD.
\]
Evaluating it at $\bx$ yields
\[
R(\bp^*)-R(\bx)
\leq\frac{\mu^2}{4(1-\mu)}\bx^\top\bD\bx
=\frac{\mu^2}{4(1-\mu)}R(\bx).
\]
Thus
\[
\frac{R(\bx)}{R(\bp^*)}
\geq
\left(1+\frac{\mu^2}{4(1-\mu)}\right)^{-1}
=\frac{4(1-\mu)}{(2-\mu)^2}.
\]
This holds for every nonzero $\ba$. Taking the infimum and using
$\mathrm{PoA}_{\min}(\bB)=\lambda_{\min}(\bM)$ proves the theorem.}

\section{Missing Proofs in \Cref{sec:Tightness}}
\subsection{Proof of \Cref{lem:symmetric-B-eigenvalues}}

From \eqref{eq:symmetric-B-matrix}, we have that $\mathbf{B} = -(1 + \rho)\mathbf{I} + \rho \mathbf{1}\mathbf{1}^\top$.
The matrix $\mathbf{1}\mathbf{1}^\top$ is a rank-one matrix with eigenvalue $N$ (with eigenvector $\mathbf{1}$) and eigenvalue $0$ with multiplicity $N - 1$ (for any vector orthogonal to $\mathbf{1}$). Then, we have that
\[
    \mathbf{B} \mathbf{1} = \left( -(1 + \rho)\mathbf{I} + \rho \mathbf{1}\mathbf{1}^\top \right) \mathbf{1} = -(1 + \rho)\mathbf{1} + \rho \mathbf{1} (\mathbf{1}^\top \mathbf{1}) = -(1 + \rho)\mathbf{1} + \rho N \mathbf{1}.
\]
Thus, it holds that
\[
    \mathbf{B} \mathbf{1} = \left( -(1 + \rho) + N\rho \right) \mathbf{1} = \left( -1 - \rho + N\rho \right) \mathbf{1} = \left( -1 + (N-1)\rho \right) \mathbf{1}.
\]
So we know that $\lambda_1 = -1 + (N-1)\rho$ with eigenvector $\mathbf{1}$.

Let $\mathbf{v}$ satisfy $\mathbf{1}^\top \mathbf{v} = 0$.
Then we have $\mathbf{1}\mathbf{1}^\top \mathbf{v} = \mathbf{1} (\mathbf{1}^\top \mathbf{v}) = \mathbf{0}$, which leads to
\[
    \mathbf{B} \mathbf{v} = -(1 + \rho) \mathbf{v} + \rho \mathbf{1}\mathbf{1}^\top \mathbf{v} = -(1 + \rho) \mathbf{v}.
\]
So we have that $\lambda_2 = -(1 + \rho) = -1 - \rho$ with multiplicity $N - 1$.

Since $\mathbf{B} = -(1+\rho)\mathbf{I} + \rho \mathbf{1}\mathbf{1}^\top$ is a rank-one perturbation of a scaled identity matrix, its inverse can be computed using the Sherman-Morrison formula (see, e.g., Section 2 of \cite{golub2013matrix}).
Since $\mathbf{B}$ has the form $\alpha \mathbf{I} + \beta \mathbf{1}\mathbf{1}^\top$ with $\alpha = -(1+\rho)$ and $\beta = \rho$, we know that its inverse has the form $\alpha' \mathbf{I} + \beta' \mathbf{1}\mathbf{1}^\top$ for some $\alpha', \beta'$.

Now that we have shown that $\mathbf{B}$ has eigenvalues $\lambda_1 = -1 + (N-1)\rho$ (for eigenvector $\mathbf{1}$) and $\lambda_2 = -1 - \rho$ (for vectors orthogonal to $\mathbf{1}$).
Therefore, $\mathbf{B}^{-1}$ has eigenvalues $1/\lambda_1$ and $1/\lambda_2$ for the same eigenvectors.
For a matrix of the form $\alpha' \mathbf{I} + \beta' \mathbf{1}\mathbf{1}^\top$, we know that the eigenvalue for vectors orthogonal to $\mathbf{1}$ is $\alpha'$, and the eigenvalue for $\mathbf{1}$ is $\alpha' + \beta' N$.
The explicit matching eigenvalues can be computed as
\[
    \alpha' = \frac{1}{\lambda_2} = \frac{1}{-1 - \rho} = -\frac{1}{1 + \rho},
\]
and
\[
    \alpha' + \beta' N = \frac{1}{\lambda_1} = \frac{1}{-1 + (N-1)\rho} = -\frac{1}{1 - (N-1)\rho}.
\]
Solving for $\beta'$, we have that
\begin{align*}
    \beta' N 
    &= \frac{-N\rho}{(1 - (N-1)\rho)(1 + \rho)}.
\end{align*}
which leads to
\[
    \beta' = \frac{-\rho}{(1 - (N-1)\rho)(1 + \rho)}.
\]
Therefore, we know that
\[
    \mathbf{B}^{-1} = -\frac{1}{1 + \rho} \mathbf{I} - \frac{\rho}{(1 + \rho)(1 - (N-1)\rho)} \mathbf{1}\mathbf{1}^\top.
\]
Our proof is thus completed.

\subsection{Proof of \Cref{prop:symmetric-optimal}}

From \Cref{thm:central-opt}, we have that $\mathbf{p}^* = -\frac{1}{2} \mathbf{B}^{-1} \mathbf{a}$.
Using \Cref{lem:symmetric-B-eigenvalues} with $\mathbf{a} = a\mathbf{1}$, we have that
\begin{align*}
    \mathbf{B}^{-1} \mathbf{a} &= \left( -\frac{1}{1 + \rho} \mathbf{I} - \frac{\rho}{(1 + \rho)(1 - \mu)} \mathbf{1}\mathbf{1}^\top \right) a \mathbf{1} = -\frac{a}{1 + \rho} \mathbf{1} - \frac{a \rho}{(1 + \rho)(1 - \mu)} \mathbf{1} (\mathbf{1}^\top \mathbf{1}) \\
    &= -\frac{a}{1 + \rho} \mathbf{1} - \frac{a \rho N}{(1 + \rho)(1 - \mu)} \mathbf{1} = -\frac{a}{1 + \rho} \left( 1 + \frac{\rho N}{1 - \mu} \right) \mathbf{1}.
\end{align*}
We simplify the term in parentheses.
Since $\mu = (N-1)\rho$, we have
\begin{align*}
    1 + \frac{\rho N}{1 - \mu} = \frac{(1 - \mu) + \rho N}{1 - \mu} = \frac{1 - (N-1)\rho + N\rho}{1 - \mu} = \frac{1 + \rho}{1 - \mu}.
\end{align*}
Therefore, it holds that
\[
    \mathbf{B}^{-1} \mathbf{a} = -\frac{a}{1 + \rho} \cdot \frac{1 + \rho}{1 - \mu} \mathbf{1} = -\frac{a}{1 - \mu} \mathbf{1},
\]
which leads to
\[
    \mathbf{p}^* = -\frac{1}{2} \mathbf{B}^{-1} \mathbf{a} = \frac{a}{2(1 - \mu)} \mathbf{1} = p^* \mathbf{1}.
\]
From \Cref{thm:central-opt}, we have $R(\mathbf{p}^*) = -\frac{1}{4} \mathbf{a}^\top \mathbf{B}^{-1} \mathbf{a}$.
We computed $\mathbf{B}^{-1} \mathbf{a} = -\frac{a}{1 - \mu} \mathbf{1}$, which leads to
\[
    \mathbf{a}^\top \mathbf{B}^{-1} \mathbf{a} = (a \mathbf{1})^\top \left( -\frac{a}{1 - \mu} \mathbf{1} \right) = -\frac{a^2}{1 - \mu} (\mathbf{1}^\top \mathbf{1}) = -\frac{N a^2}{1 - \mu}.
\]
Therefore, we have that
\[
    R(\mathbf{p}^*) = -\frac{1}{4} \left( -\frac{N a^2}{1 - \mu} \right) = \frac{N a^2}{4(1 - \mu)}.
\]
Our proof is thus completed.

{\subsection{Proof of \Cref{prop:symmetric-nash}}

In this model, $b_{ii}=-1$ and $b_{ij}=\rho$ for $i\ne j$, so
\[
\bA^{\mathrm{NE}}
=-(2+\rho)\bI+\rho\mathbf1\mathbf1^\top,
\qquad \mu=(N-1)\rho<1.
\]
We compute its action on the two orthogonal invariant subspaces
explicitly. Since $\mathbf1^\top\mathbf1=N$,
\begin{align*}
\bA^{\mathrm{NE}}\mathbf1
&=-(2+\rho)\mathbf1+\rho\mathbf1(\mathbf1^\top\mathbf1)\\
&=\bigl(-2+(N-1)\rho\bigr)\mathbf1
=-(2-\mu)\mathbf1.
\end{align*}
For every $\bv$ with $\mathbf1^\top\bv=0$,
\[
\bA^{\mathrm{NE}}\bv
=-(2+\rho)\bv+\rho\mathbf1(\mathbf1^\top\bv)
=-(2+\rho)\bv.
\]
These subspaces span $\R^N$. Both eigenvalues are strictly negative,
so $\bA^{\mathrm{NE}}$ is invertible. The unique solution of
$\bA^{\mathrm{NE}}\bx=-a\mathbf1$ is therefore
\[
\bx=\frac{a}{2-\mu}\mathbf1.
\]
For completeness, this solution is a Nash equilibrium because, for
fixed opponents' prices, each payoff has second derivative
$2b_{ii}=-2<0$. Each coordinate of the displayed solution satisfies
that player's first-order condition and is its unique best response.

At these prices, every product has demand
\begin{align*}
F_i(\bx)
&=a-\frac{a}{2-\mu}
   +\rho(N-1)\frac{a}{2-\mu}\\
&=\frac{a}{2-\mu}.
\end{align*}
Summing the product revenues gives
\[
R(\bx)=\sum_{i=1}^N x_iF_i(\bx)
      =\frac{Na^2}{(2-\mu)^2},
\]
which proves both asserted formulas.}

\subsection{Proof of \Cref{thm:PoA-tight}}

From \Cref{prop:symmetric-optimal} and \Cref{prop:symmetric-nash}, we have that
\[
    R(\mathbf{p}^*) = \frac{N a^2}{4(1 - \mu)}, \qquad R(\mathbf{p}^{\mathrm{NE}}) = \frac{N a^2}{(2 - \mu)^2}.
\]
Therefore, it holds that
\[
    \mathrm{PoA}(\mathbf{a}) = \frac{R(\mathbf{p}^{\mathrm{NE}})}{R(\mathbf{p}^*)} = \frac{\frac{N a^2}{(2 - \mu)^2}}{\frac{N a^2}{4(1 - \mu)}} = \frac{N a^2}{(2 - \mu)^2} \cdot \frac{4(1 - \mu)}{N a^2} = \frac{4(1 - \mu)}{(2 - \mu)^2}.
\]
This exactly equals the bound from \Cref{thm:PoA-mu-bound}.

\section{Missing Proofs in \Cref{sec:spectral}}

{\subsection{Proof of \Cref{lem:M-norm-properties}}

Define
\[
\widehat\mu:=\max_i\mu_i,
\qquad
\mu_i=\frac1{-b_{ii}}
\max\left\{\sum_{j\ne i}|b_{ij}|,
           \sum_{j\ne i}|b_{ji}|\right\}.
\]
The standing assumptions imply $0\leq\widehat\mu\leq\mu<1$.
Let $\mathbf C=\bB+\bD$, so that
$\bM_{\mathrm{norm}}=\bD^{-1/2}\mathbf C\bD^{-1/2}$.
The definition of $\widehat\mu$ gives both
$\sum_j|C_{ij}|\leq\widehat\mu(-b_{ii})$ and
$\sum_j|C_{ji}|\leq\widehat\mu(-b_{ii})$.
For any $\bv\in\R^N$, weighted Cauchy-Schwarz gives
\begin{align*}
\sum_i\frac{|(\mathbf C\bv)_i|^2}{-b_{ii}}
\leq\sum_i\frac{\sum_j|C_{ij}|}{-b_{ii}}
                  \sum_j|C_{ij}|v_j^2\leq\widehat\mu\sum_jv_j^2\sum_i|C_{ij}|\leq\widehat\mu^2\sum_j(-b_{jj})v_j^2.
\end{align*}
The first inequality follows by applying Cauchy--Schwarz to
$\sum_j\sqrt{|C_{ij}|}\,\bigl(\sqrt{|C_{ij}|}|v_j|\bigr)$;
the next two inequalities use the row and column bounds,
respectively. With $\bv=\bD^{-1/2}\bz$, this becomes
\[
\|\bM_{\mathrm{norm}}\bz\|_2^2
\leq\widehat\mu^2\|\bz\|_2^2.
\]
Taking the supremum over unit vectors $\bz$ proves
\[
0\leq\mu_{\mathrm{spectral}}
=\|\bM_{\mathrm{norm}}\|_2
\leq\widehat\mu\leq\mu<1.
\]

Finally,
$\mathbf S=(\bM_{\mathrm{norm}}+\bM_{\mathrm{norm}}^\top)/2$
is symmetric and satisfies
\[
\|\mathbf S\|_2
\leq\frac12\left(\|\bM_{\mathrm{norm}}\|_2
                 +\|\bM_{\mathrm{norm}}^\top\|_2\right)
=\mu_{\mathrm{spectral}}.
\]
If $\mathbf S\bv=\lambda\bv$ with $\bv\ne\mathbf0$, then
$|\lambda|\|\bv\|_2=\|\mathbf S\bv\|_2
\leq\mu_{\mathrm{spectral}}\|\bv\|_2$.
Thus every eigenvalue of $\mathbf S$ lies in the asserted interval.}

{\subsection{Proof of \Cref{lem:revenue-gap}}

Write
\[
\bH=-\bB_{\mathrm{s}},\quad
\bG=\bD-\bB,\quad
\mathbf C=\bB+\bD,\quad
\bx=\bG^{-1}\ba.
\]
The assumptions give $\bH\succ0$ and $\bD\succ0$.
The equilibrium equation $\ba=(\bD-\bB)\bx$ implies
\[
\bm F(\bx)=\ba+\bB\bx=\bD\bx,
\qquad R(\bx)=\bx^\top\bD\bx.
\]
Also, it holds
\[
\ba-2\bH\bx
=(\bD-\bB+\bB+\bB^\top)\bx
=\mathbf C^\top\bx.
\]
Let $\boldsymbol\Delta=\bp-\bx$ for an arbitrary $\bp\in\R^N$.
Expanding the quadratic revenue function gives
\begin{align*}
R(\bp)-R(\bx)
&=\ba^\top\boldsymbol\Delta
 -2\bx^\top\bH\boldsymbol\Delta
 -\boldsymbol\Delta^\top\bH\boldsymbol\Delta=(\mathbf C^\top\bx)^\top\boldsymbol\Delta
 -\boldsymbol\Delta^\top\bH\boldsymbol\Delta\\
&=\frac14\bx^\top\mathbf C\bH^{-1}\mathbf C^\top\bx-
 \left(\boldsymbol\Delta-\frac12\bH^{-1}\mathbf C^\top\bx\right)^\top
 \bH
 \left(\boldsymbol\Delta-\frac12\bH^{-1}\mathbf C^\top\bx\right).
\end{align*}
The last term is nonnegative and is zero at
$\boldsymbol\Delta_*=\frac12\bH^{-1}\mathbf C^\top\bx$.
Because $\bp$ ranges over all of $\R^N$, this displacement is feasible.
Indeed, we have
\[
\bx+\boldsymbol\Delta_*
=\frac12\bH^{-1}(2\bH+\mathbf C^\top)\bx
=\frac12\bH^{-1}\ba=\bp^*.
\]
Taking the maximum over $\bp$ proves the exact identity
\[
R(\bp^*)-R(\bx)
=\frac14\bx^\top\mathbf C\bH^{-1}\mathbf C^\top\bx.
\]
To express this in normalized coordinates, use
\[
\mathbf C=\bD^{1/2}\bM_{\mathrm{norm}}\bD^{1/2},
\qquad
\bH=\bD^{1/2}(\bI-\mathbf S)\bD^{1/2}.
\]
Inverting the second factorization and substituting both gives
\begin{align*}
\mathbf C\bH^{-1}\mathbf C^\top
&=\bD^{1/2}\bM_{\mathrm{norm}}
   (\bI-\mathbf S)^{-1}\bM_{\mathrm{norm}}^\top\bD^{1/2}=\bD^{1/2}\bQ\bD^{1/2}.
\end{align*}
If $\ba\ne\mathbf0$, then $\bx\ne\mathbf0$ and
$\bz=\bD^{1/2}\bx\ne\mathbf0$. Therefore $R(\bx)=\bz^\top\bz>0$,
and the gap identity becomes
\[
R(\bp^*)=\bz^\top\bz+\frac14\bz^\top\bQ\bz.
\]
Dividing $R(\bx)$ by this expression yields
\[
\mathrm{PoA}(\ba)
=\left(1+\frac{\bz^\top\bQ\bz}{4\bz^\top\bz}\right)^{-1},
\]
as claimed.}

\subsection{Proof of \Cref{thm:PoA-spectral-exact}}

Let
\[
\bQ=\bM_{\mathrm{norm}}(\bI-\mathbf S)^{-1}
                         \bM_{\mathrm{norm}}^\top,
\qquad
\bz=\bD^{1/2}\bG^{-1}\ba.
\]
The matrix $\bI-\mathbf S$ is positive definite, so $\bQ$ is
symmetric positive semidefinite. The revenue-gap identity in
\Cref{lem:revenue-gap} states that
\[
R(\bp^{\mathrm{NE}})=\bz^\top\bz,
\qquad
R(\bp^*)=\bz^\top\bz+\frac14\bz^\top\bQ\bz.
\]
Consequently, we have
\[
\mathrm{PoA}(\ba)
=\left(1+\frac14
  \frac{\bz^\top\bQ\bz}{\bz^\top\bz}\right)^{-1}.
\]
The map $\ba\mapsto\bz$ is a bijection on nonzero real vectors,
with inverse $\ba=\bG\bD^{-1/2}\bz$.

Let $\bv_1,\ldots,\bv_N$ be an orthonormal eigenbasis of $\bQ$,
with eigenvalues $q_1,\ldots,q_N\geq0$. For
$\bz=\sum_i\alpha_i\bv_i\ne\mathbf0$,
\[
\frac{\bz^\top\bQ\bz}{\bz^\top\bz}
=\frac{\sum_i q_i\alpha_i^2}{\sum_i\alpha_i^2}
\leq\max_i q_i=\lambda_{\max}(\bQ).
\]
Equality holds when $\bz$ is a nonzero eigenvector associated with
$\lambda_{\max}(\bQ)$. Since $t\mapsto(1+t/4)^{-1}$ is decreasing
for $t\geq0$, this proves
\[
\mathrm{PoA}_{\min}(\bB)
=\left(1+\frac14\lambda_{\max}(\bQ)\right)^{-1}.
\]
An attaining vector $\bz=\bv^*$ corresponds exactly to
$\ba^*=\bG\bD^{-1/2}\bv^*\ne\mathbf0$.

Define $\mathbf W=\bM_{\mathrm{norm}}(\bI-\mathbf S)^{-1/2}$.
Then we have $\bQ=\mathbf W\mathbf W^\top$, and
\[
\lambda_{\max}(\bQ)
=\max_{\|\bv\|_2=1}\bv^\top\mathbf W\mathbf W^\top\bv
=\max_{\|\bv\|_2=1}\|\mathbf W^\top\bv\|_2^2
=\|\mathbf W\|_2^2.
\]
This establishes the singular-value expression in \eqref{eq:PoA-exact}.

For the scalar lower bounds, write
$r=\mu_{\mathrm{spectral}}=\|\bM_{\mathrm{norm}}\|_2$.
The hypotheses of \Cref{lem:M-norm-properties} are the standing
assumptions, so that lemma gives $0\leq r\leq\mu<1$ and
$\|\mathbf S\|_2\leq r$. Diagonalizing $\mathbf S$ shows
\[
(\bI-\mathbf S)^{-1}\preceq\frac1{1-r}\bI.
\]
Thus, for every $\bv$, we have
\begin{align*}
\bv^\top\bQ\bv
&=(\bM_{\mathrm{norm}}^\top\bv)^\top
       (\bI-\mathbf S)^{-1}(\bM_{\mathrm{norm}}^\top\bv)\leq\frac1{1-r}\|\bM_{\mathrm{norm}}^\top\bv\|_2^2\leq\frac{r^2}{1-r}\|\bv\|_2^2.
\end{align*}
It follows that $\lambda_{\max}(\bQ)\leq r^2/(1-r)$ and hence it holds
\[
\mathrm{PoA}_{\min}(\bB)
\geq\left(1+\frac{r^2}{4(1-r)}\right)^{-1}
=\frac{4(1-r)}{(2-r)^2}.
\]
For $f(t)=4(1-t)/(2-t)^2$, differentiation gives
$f'(t)=-4t/(2-t)^3\leq0$ on $[0,1)$.
Since $r\leq\mu$, we obtain $f(r)\geq f(\mu)$, proving
\eqref{eq:PoA-spectral-norm-bound}. The case $r=0$ is included:
then $\bM_{\mathrm{norm}}=\bQ=0$ and the ratio is one.

Finally, we verify the symmetric specialization stated after the
theorem. Suppose $\bM_{\mathrm{norm}}$ happens to be symmetric and
write $\bM_{\mathrm{norm}}\bv_i=\lambda_i\bv_i$ in an orthonormal
basis. Then $\mathbf S=\bM_{\mathrm{norm}}$ and
\[
\bQ\bv_i
=\frac{\lambda_i^2}{1-\lambda_i}\bv_i,
\qquad -1<\lambda_i<1.
\]
Substituting these eigenvalues into the exact formula gives
\[
\mathrm{PoA}_{\min}(\bB)
=\min_i\left(1+\frac{\lambda_i^2}{4(1-\lambda_i)}\right)^{-1}
=\min_i f(\lambda_i).
\]
The derivative of $f$ is positive on $(-1,0)$ and negative on
$(0,1)$; hence the minimum over a finite spectrum is attained at
its smallest or largest eigenvalue. In this specialization
$r=\max_i|\lambda_i|$. If $r>0$, direct subtraction yields
\[
f(r)-f(-r)=-\frac{8r^3}{(4-r^2)^2}<0.
\]
For $\lambda\in[-r,0]$, $f(\lambda)\geq f(-r)>f(r)$.
For $\lambda\in[0,r)$, strict monotonicity gives $f(\lambda)>f(r)$.
Therefore $\min_i f(\lambda_i)=f(r)$ if and only if $r$ is an
eigenvalue, equivalently $\lambda_{\max}=r$.
When $r=0$, all eigenvalues are zero and the same equivalence holds.

{\section{Missing Proofs in \Cref{sec:nonlinear}}
\label{app:nonlinear}

{\subsection{Proof of \Cref{prop:nl-equilibrium}}

Fix $\bp\in\mathcal P$. By definition,
$H_{ii}(\bp)=-b_{ii}(\bp)>0$ and, for $i\ne j$,
$H_{ij}(\bp)=-[b_{ij}(\bp)+b_{ji}(\bp)]/2$.
Two-sided Jacobian dominance implies
\[
\sum_{j\ne i}|H_{ij}(\bp)|
\leq\frac12\left(\sum_{j\ne i}|b_{ij}(\bp)|
                +\sum_{j\ne i}|b_{ji}(\bp)|\right)
\leq\mu[-b_{ii}(\bp)].
\]
For any $\bv\in\R^N$, symmetry of $\bH(\bp)$ and
$2|v_iv_j|\leq v_i^2+v_j^2$ give
\begin{align*}
\left|\bv^\top[\bH(\bp)-\bD(\bp)]\bv\right|
\leq\frac12\sum_{i\ne j}|H_{ij}(\bp)|(v_i^2+v_j^2)=\sum_i v_i^2\sum_{j\ne i}|H_{ij}(\bp)|\leq\mu\bv^\top\bD(\bp)\bv.
\end{align*}
In particular,
$\bH(\bp)\succeq(1-\mu)\bD(\bp)\succ0$.
Differentiating $R(\bp)=\sum_i p_iF_i(\bp)$ twice yields
\[
\nabla^2R(\bp)
=\bB(\bp)+\bB(\bp)^\top
 +\sum_i p_i\nabla^2F_i(\bp).
\]
Each $p_i$ is nonnegative and each demand Hessian is negative
semidefinite. Hence, for every nonzero $\bv$,
\[
\bv^\top\nabla^2R(\bp)\bv
\leq-2(1-\mu)\bv^\top\bD(\bp)\bv<0.
\]
This proves strict concavity of $R$ on the convex box $\mathcal P$.
Continuity and compactness give existence of a maximizer, and
strict concavity gives uniqueness.

Fix player $i$ and the opponents' prices. The own-price second
derivative is
\[
\frac{\partial^2u_i(\bp)}{\partial p_i^2}
=2b_{ii}(\bp)+p_i\partial_i b_{ii}(\bp)<0.
\]
Indeed, $b_{ii}(\bp)<0$, while joint demand concavity implies
$\partial_i b_{ii}(\bp)=\partial_{ii}F_i(\bp)\leq0$.
The continuous own payoff therefore has a unique maximizer on
$[0,\overline p_i]$; denote it by $T_i(\bp_{-i})$.

We verify continuity of this best response. If
$\bp_{-i}^{(n)}\to\bp_{-i}$, let
$t_n=T_i(\bp_{-i}^{(n)})$. Any subsequence of $(t_n)$ has a further
subsequence converging to some $t\in[0,\overline p_i]$.
For every $q_i\in[0,\overline p_i]$, optimality gives
\[
u_i(t_n,\bp_{-i}^{(n)})\geq u_i(q_i,\bp_{-i}^{(n)}).
\]
Passing to the limit along that further subsequence yields
$u_i(t,\bp_{-i})\geq u_i(q_i,\bp_{-i})$ for every $q_i$.
Uniqueness of the maximizer implies $t=T_i(\bp_{-i})$.
Every subsequential limit has this value, so the full sequence
converges to it. Thus $T_i$ is continuous.

The simultaneous best-response map
$T(\bp)=(T_i(\bp_{-i}))_{i=1}^N$ is a continuous self-map of the
nonempty compact convex set $\mathcal P$.
Brouwer's fixed-point theorem \citep{boothby1971two} gives a fixed
point. At that point every player is playing its best response,
so it is a pure-strategy Nash equilibrium.

For fixed opponents, differentiable concavity implies that $x_i$
is optimal if and only if
\[
g_i(\bx)(p_i-x_i)\leq0
\qquad\text{for every }p_i\in[0,\overline p_i].
\]
Necessity follows by differentiating the payoff along the feasible
segment from $x_i$ to $p_i$ at its maximizing endpoint.
For sufficiency, the concavity inequality gives
\[
u_i(p_i,\bx_{-i})-u_i(x_i,\bx_{-i})
\leq g_i(\bx)(p_i-x_i)\leq0.
\]
Summing over $i$ gives \eqref{eq:nl-equilibrium-vi}. Conversely,
varying only coordinate $i$ in that vector inequality recovers
its scalar counterpart, and therefore the best-response condition.
This reasoning includes both endpoints of every price interval.

Taking $\bp=\mathbf0$ in the vector inequality gives
$\bx^\top\mathbf g(\bx)\geq0$. Since
$\bm F(\bx)=\bD(\bx)\bx+\mathbf g(\bx)$,
\[
R(\bx)=\bx^\top\bm F(\bx)
=\bx^\top\bD(\bx)\bx+\bx^\top\mathbf g(\bx)
\geq\bx^\top\bD(\bx)\bx.
\]
If $\bx$ is interior, both signs of a sufficiently small coordinate
deviation are feasible, forcing $g_i(\bx)=0$ for every $i$.
The revenue inequality is then an equality.}

\subsection{Proof of \Cref{lem:nl-tangent}}

Fix an equilibrium $\bx$ and suppress the argument $\bx$ in
$\bB,\bD,\bH,\mathbf C$, and $\mathbf g$. For
$\boldsymbol\Delta=\bp-\bx$, define
\[
r_i:=F_i(\bp)-F_i(\bx)-\nabla F_i(\bx)^\top\boldsymbol\Delta,
\qquad \mathbf r=(r_1,\ldots,r_N)^\top.
\]
The vector identity
$\bm F(\bp)=\bm F(\bx)+\bB\boldsymbol\Delta+\mathbf r$
gives the exact expansion
\[
R(\bp)-R(\bx)
=\boldsymbol\Delta^\top\bm F(\bx)
  +\bx^\top\bB\boldsymbol\Delta
  +\boldsymbol\Delta^\top\bB\boldsymbol\Delta
  +\sum_i p_i r_i\notag=\mathbf g^\top\boldsymbol\Delta
  +(\mathbf C^\top\bx)^\top\boldsymbol\Delta
  -\boldsymbol\Delta^\top\bH\boldsymbol\Delta
  +\sum_i p_i r_i.
\]
Here we used $\bm F(\bx)=\bD\bx+\mathbf g$,
$\bD+\bB^\top=\mathbf C^\top$, and
$\boldsymbol\Delta^\top\bB\boldsymbol\Delta
=-\boldsymbol\Delta^\top\bH\boldsymbol\Delta$.

Joint concavity implies $r_i\leq0$, so $\sum_i p_i r_i\leq0$
because prices are nonnegative. The equilibrium inequality gives
$\mathbf g^\top\boldsymbol\Delta\leq0$. Removing these two
nonpositive terms from the above equation proves
\eqref{eq:nl-tangent-bound}.

{\subsection{Proof of \Cref{thm:nl-poa}}

Fix a Nash equilibrium $\bx\in\mathcal P$. Throughout this proof,
all matrices without an argument are evaluated at $\bx$:
\[
\bB=\bB(\bx),\quad
\bD=\bD(\bx),\quad
\bH=-\frac{\bB+\bB^\top}{2},\quad
\mathbf C=\bB+\bD,\quad
\mathbf g=\bm F(\bx)-\bD\bx.
\]
In particular, $b_{ii}=b_{ii}(\bx)<0$ in the coordinate calculations
below. Put
\[
\mathbf T=\bD^{-1/2}\mathbf C\bD^{-1/2}
          =\bM_{\mathrm{norm}}(\bx),
\qquad
\mathbf S=\frac{\mathbf T+\mathbf T^\top}{2}.
\]

Because $\bx$ maximizes each player's payoff against its fixed
opponents, differentiation along any feasible coordinate segment
gives $g_i(p_i-x_i)\leq0$ for every $p_i\in[0,\overline p_i]$.
Summing, and then choosing $\bp=\mathbf0$, gives
\[
\mathbf g^\top(\bp-\bx)\leq0,
\qquad
\bx^\top\mathbf g\geq0,
\qquad
R(\bx)=\bx^\top\bD\bx+\bx^\top\mathbf g
       \geq\bx^\top\bD\bx.
\]
No interiority assumption is used.

For any $\bp\in\mathcal P$, let $\boldsymbol\Delta=\bp-\bx$ and
\[
r_i=F_i(\bp)-F_i(\bx)
                 -\nabla F_i(\bx)^\top\boldsymbol\Delta.
\]
Joint demand concavity implies $r_i\leq0$. Expanding
$\sum_i p_i[F_i(\bx)+\nabla F_i(\bx)^\top\boldsymbol\Delta+r_i]$
and subtracting $R(\bx)$ gives
\begin{align*}
R(\bp)-R(\bx)
&=\boldsymbol\Delta^\top\bm F(\bx)
 +\bx^\top\bB\boldsymbol\Delta
 +\boldsymbol\Delta^\top\bB\boldsymbol\Delta
 +\sum_i p_i r_i\\
&=\mathbf g^\top\boldsymbol\Delta
 +(\mathbf C^\top\bx)^\top\boldsymbol\Delta
 -\boldsymbol\Delta^\top\bH\boldsymbol\Delta
 +\sum_i p_i r_i\\
&\leq(\mathbf C^\top\bx)^\top\boldsymbol\Delta
 -\boldsymbol\Delta^\top\bH\boldsymbol\Delta.
\end{align*}
The last inequality uses the equilibrium condition and $p_i\geq0$.

Pointwise two-sided dominance gives
$\sum_j|C_{ij}|\leq\mu(-b_{ii})$ and
$\sum_j|C_{ji}|\leq\mu(-b_{ii})$, since $C_{ii}=0$.
For any $\bv\in\R^N$, weighted Cauchy--Schwarz therefore yields
\begin{align*}
\sum_i\frac{|(\mathbf C\bv)_i|^2}{-b_{ii}}\leq\sum_i\frac{\sum_j|C_{ij}|}{-b_{ii}}
                    \sum_j|C_{ij}|v_j^2\leq\mu\sum_jv_j^2\sum_i|C_{ij}|\leq\mu^2\sum_j(-b_{jj})v_j^2.
\end{align*}
Substituting $\bv=\bD^{-1/2}\bz$ proves
$\|\mathbf T\|_2\leq\mu$. Write
$r=\|\mathbf T\|_2=\mu_{\mathrm{spectral}}(\bx)<1$.
Then $\|\mathbf S\|_2\leq r$ and
\[
\bH=\bD^{1/2}(\bI-\mathbf S)\bD^{1/2}\succ0,
\qquad
(\bI-\mathbf S)^{-1}\preceq\frac1{1-r}\bI.
\]
These conclusions follow by diagonalizing the symmetric matrix
$\mathbf S$, whose eigenvalues have absolute values at most $r$.

For $\mathbf e=\mathbf C^\top\bx$, completing the square gives
\begin{align*}
\mathbf e^\top\boldsymbol\Delta
 -\boldsymbol\Delta^\top\bH\boldsymbol\Delta
=\frac14\mathbf e^\top\bH^{-1}\mathbf e\quad-
 \left(\boldsymbol\Delta-\frac12\bH^{-1}\mathbf e\right)^\top
 \bH\left(\boldsymbol\Delta-\frac12\bH^{-1}\mathbf e\right)\leq\frac14\bx^\top\mathbf C\bH^{-1}\mathbf C^\top\bx.
\end{align*}
Apply the tangent comparison at a centralized maximizer to obtain
\eqref{eq:nl-gap-certificate}. Such a maximizer exists because
$R$ is continuous on the compact box. The displacement maximizing
the quadratic upper bound need not be feasible; only its upper-bound
value is used.

Define $\bQ=\mathbf T(\bI-\mathbf S)^{-1}\mathbf T^\top=\bQ(\bx)$.
For every $\bv$, we have
\begin{align*}
\bv^\top\bQ\bv
=(\mathbf T^\top\bv)^\top
        (\bI-\mathbf S)^{-1}(\mathbf T^\top\bv)\leq\frac1{1-r}\|\mathbf T^\top\bv\|_2^2
 \leq\frac{r^2}{1-r}\|\bv\|_2^2.
\end{align*}
Thus $\lambda_{\max}(\bQ)\leq r^2/(1-r)\leq\mu^2/(1-\mu)$.
The last inequality follows because $t^2/(1-t)$ is nondecreasing
on $[0,1)$.
Using the factorizations of $\mathbf C$ and $\bH$, we also have
\[
\mathbf C\bH^{-1}\mathbf C^\top
=\bD^{1/2}\bQ\bD^{1/2}.
\]
Consequently, it holds
\begin{align*}
R^*
\leq R(\bx)+\frac14\bx^\top\mathbf C\bH^{-1}\mathbf C^\top\bx\leq R(\bx)+\frac{\mu^2}{4(1-\mu)}\bx^\top\bD\bx\leq\left(1+\frac{\mu^2}{4(1-\mu)}\right)R(\bx).
\end{align*}
Since $R^*>0$, this proves $R(\bx)>0$ before any division by
$R(\bx)$.

Let $\bz=\bD^{1/2}\bx$. Then
$\|\bz\|_2^2=\bx^\top\bD\bx\leq R(\bx)$ and
\[
\bx^\top\mathbf C\bH^{-1}\mathbf C^\top\bx
=\bz^\top\bQ\bz
\leq\lambda_{\max}(\bQ)R(\bx).
\]
Combining this with the gap bound gives, in order,
\begin{align*}
\frac{R(\bx)}{R^*}
\geq
 \left(1+\frac{\bx^\top\mathbf C\bH^{-1}\mathbf C^\top\bx}
                  {4R(\bx)}\right)^{-1}\geq\left(1+\frac14\lambda_{\max}(\bQ)\right)^{-1}\geq\left(1+\frac{r^2}{4(1-r)}\right)^{-1}
 =\frac{4(1-r)}{(2-r)^2}\geq\frac{4(1-\mu)}{(2-\mu)^2}.
\end{align*}
The last step uses the derivative
$\frac{d}{dt}[4(1-t)/(2-t)^2]=-4t/(2-t)^3\leq0$.
Restoring the arguments $\bx$ proves \eqref{eq:nl-poa-chain}.
The equilibrium was arbitrary, so taking the infimum over
$\mathcal E(\bm F,\mathcal P)$ proves the asserted PoA guarantee.}

{\subsection{Proof of \Cref{prop:nl-uniqueness}}

For fixed opponents' prices, each payoff is continuous on
$[0,\overline p_i]$ and has second derivative
$2b_{ii}(\bp)+p_i\partial_i b_{ii}(\bp)<0$.
Thus its maximizer exists and is unique; denote it by
$T_i(\bp_{-i})$. The simultaneous best-response map is
$T(\bp)=(T_i(\bp_{-i}))_{i=1}^N$ and maps $\mathcal P$ into itself.
If $N=1$, this map is constant, so the conclusions are immediate.
For $N\geq2$, fix $\bp,\mathbf q\in\mathcal P$ and a player $i$,
and put
\[
u=T_i(\bp_{-i}),\qquad v=T_i(\mathbf q_{-i}),\qquad
h=\|\bp_{-i}-\mathbf q_{-i}\|_\infty.
\]

Suppose $u>v$. Since $u>0$, the price $u$ is either interior or
at the upper endpoint. Optimality therefore gives
$g_i(u,\bp_{-i})\geq0$. Since $v<\overline p_i$, the price $v$
is either interior or at the lower endpoint, and
$g_i(v,\mathbf q_{-i})\leq0$.
In particular,
\[
g_i(u,\bp_{-i})-g_i(v,\mathbf q_{-i})\geq0.
\]
The entire segment
\[
\bp(t)=\bigl(v+t(u-v),\ \mathbf q_{-i}
                          +t(\bp_{-i}-\mathbf q_{-i})\bigr),
\qquad 0\leq t\leq1,
\]
lies in $\mathcal P$. The chain rule and the marginal-revenue
condition give
\begin{align*}
\frac{d}{dt}g_i(\bp(t))
&=\partial_i g_i(\bp(t))(u-v)
  +\sum_{j\ne i}\partial_jg_i(\bp(t))(p_j-q_j)\\
&\leq\partial_i g_i(\bp(t))(u-v)
       +h\sum_{j\ne i}|\partial_jg_i(\bp(t))|\\
&\leq[-\partial_i g_i(\bp(t))]\,[\kappa h-(u-v)].
\end{align*}
Here $-\partial_i g_i(\bp(t))>0$, and
$\partial_jg_i=b_{ij}+p_i\partial_i b_{ij}$ for $j\ne i$,
where equality of mixed partials justifies the latter identity.
If $u-v>\kappa h$, integration would imply
\[
g_i(u,\bp_{-i})-g_i(v,\mathbf q_{-i})
\leq[\kappa h-(u-v)]
       \int_0^1[-\partial_i g_i(\bp(t))]\,dt<0,
\]
contradicting the nonnegative difference above.
Thus $u-v\leq\kappa h$. When $v>u$, applying this established
inequality to the ordered pair $(\mathbf q,\bp)$ gives
$v-u\leq\kappa\|\mathbf q_{-i}-\bp_{-i}\|_\infty=\kappa h$.
When $u=v$, the absolute difference is zero.
In all cases,
\[
|T_i(\bp_{-i})-T_i(\mathbf q_{-i})|
\leq\kappa\|\bp_{-i}-\mathbf q_{-i}\|_\infty.
\]
Taking the maximum over $i$ proves
\[
\|T(\bp)-T(\mathbf q)\|_\infty
\leq\kappa\|\bp-\mathbf q\|_\infty.
\]

For completeness, this estimate implies both existence and convergence
of the fixed point. If $\kappa=0$, $T$ is constant and its value is
its unique fixed point; one update reaches it.
For $0<\kappa<1$, start from any $\bp^0\in\mathcal P$ and define
$\bp^{t+1}=T(\bp^t)$. Induction gives
\[
\|\bp^{t+1}-\bp^t\|_\infty
\leq\kappa^t\|\bp^1-\bp^0\|_\infty.
\]
For integers $m>t$, the triangle inequality yields
\[
\|\bp^m-\bp^t\|_\infty
\leq\frac{\kappa^t}{1-\kappa}\|\bp^1-\bp^0\|_\infty.
\]
The iterates are therefore Cauchy and converge in the closed box to
a vector $\bx$. Continuity of $T$, which follows from its Lipschitz
estimate, gives $T(\bx)=\bx$.
If $T(\mathbf y)=\mathbf y$ as well, then
$\|\bx-\mathbf y\|_\infty\leq\kappa\|\bx-\mathbf y\|_\infty$,
forcing $\bx=\mathbf y$. Fixed points are precisely Nash equilibria,
so $\bx=\bp^{\mathrm{NE}}$ is the unique equilibrium.
Finally, repeated application of the contraction estimate gives
\[
\|\bp^t-\bp^{\mathrm{NE}}\|_\infty
\leq\kappa^t\|\bp^0-\bp^{\mathrm{NE}}\|_\infty.
\]
In particular, simultaneous best responses converge from every
initial vector in $\mathcal P$.}

{\subsection{Parameter Conditions for \Cref{ex:nl-cross}}

Consider
\[
F_i(\bp)=a_i+\sum_jb_{ij}p_j-\gamma_i(\bw^\top\bp)^2,
\qquad b_{ii}<0,\quad \gamma_i\geq0,\quad\bw\in\R_+^N.
\]
Let $W=\bw^\top\overline{\bp}$, where
$\overline{\bp}=(\overline p_1,\ldots,\overline p_N)^\top$.
For every $\bp\in\mathcal P$, $0\leq\bw^\top\bp\leq W$ and
\[
F_i(\bp)
\geq a_i-(-b_{ii})\overline p_i
       -\sum_{\substack{j\ne i:\ b_{ij}<0}}|b_{ij}|\overline p_j
       -\gamma_iW^2.
\]
Here the nonnegative affine cross terms were dropped, while each
negative affine term and the nonpositive quadratic term were
bounded below over the box. Thus the sufficient conditions
\begin{equation}
a_i\geq(-b_{ii})\overline p_i
  +\sum_{\substack{j\ne i:\ b_{ij}<0}}|b_{ij}|\overline p_j
  +\gamma_iW^2
\qquad\text{for every }i
\label{eq:nl-cross-nonnegative}
\end{equation}
ensure nonnegative demand throughout $\mathcal P$.

Direct differentiation yields
\[
b_{ij}(\bp)=b_{ij}-2\gamma_iw_j(\bw^\top\bp),
\qquad
\nabla^2F_i(\bp)=-2\gamma_i\bw\bw^\top.
\]
The functions are smooth, and for every $\bv$,
$\bv^\top\nabla^2F_i(\bp)\bv=-2\gamma_i(\bw^\top\bv)^2\leq0$.
They are therefore jointly concave. Also,
\[
-b_{ii}(\bp)=-b_{ii}+2\gamma_iw_i(\bw^\top\bp)
\geq-b_{ii}>0.
\]
For $i\ne j$, the triangle inequality gives
\[
|b_{ij}(\bp)|\leq|b_{ij}|+2\gamma_iw_jW,
\qquad
|b_{ji}(\bp)|\leq|b_{ji}|+2\gamma_jw_iW.
\]
Consequently, a sufficient condition for two-sided Jacobian dominance
is the pair of inequalities
\begin{equation*}
\begin{aligned}
\sum_{j\ne i}|b_{ij}|+2\gamma_iW\sum_{j\ne i}w_j
\leq\mu(-b_{ii}),\quad\sum_{j\ne i}|b_{ji}|+2w_iW\sum_{j\ne i}\gamma_j
\leq\mu(-b_{ii})
\qquad\text{for every }i.
\end{aligned}
\end{equation*}
Indeed, the first gives
\[
\sum_{j\ne i}|b_{ij}(\bp)|
\leq\sum_{j\ne i}|b_{ij}|+2\gamma_iW\sum_{j\ne i}w_j
\leq\mu(-b_{ii})\leq\mu[-b_{ii}(\bp)],
\]
and the second gives
\[
\sum_{j\ne i}|b_{ji}(\bp)|
\leq\sum_{j\ne i}|b_{ji}|+2w_iW\sum_{j\ne i}\gamma_j
\leq\mu(-b_{ii})\leq\mu[-b_{ii}(\bp)].
\]
Thus \eqref{eq:nl-diagdom} holds at every feasible price vector.
Together with \eqref{eq:nl-cross-nonnegative}, these conditions verify
the smoothness, nonnegativity, own-price, joint-concavity, and
Jacobian-dominance assumptions of \Cref{thm:nl-poa} for this example.
Moreover, \eqref{eq:nl-cross-nonnegative} gives
$a_i\geq(-b_{ii})\overline p_i>0$. If only price $i$ is raised
from zero to a sufficiently small $t>0$, continuity gives
$F_i(0,\ldots,t,\ldots,0)>0$. The total revenue at that vector
is $tF_i(0,\ldots,t,\ldots,0)>0$, verifying $R^*>0$ as well.}

{\subsection{Proof of \Cref{prop:nl-curvature}}

We maintain the conditions stated at the beginning of
\Cref{subsec:nl-curvature}: smooth and nonnegative demand, negative
own-price effects, two-sided Jacobian dominance, the common compact
price box, and $R^*>0$. Joint demand concavity is replaced by
\eqref{eq:nl-own-revenue-concavity}.

For each player and fixed opponents, the own payoff is continuous on
$[0,\overline p_i]$ and its second derivative is
\[
\partial_{ii}u_i(\bp)
=2b_{ii}(\bp)+p_i\partial_i b_{ii}(\bp)<0.
\]
It therefore has a unique maximizer, denoted $T_i(\bp_{-i})$.
To verify continuity, let $\bp_{-i}^{(n)}\to\bp_{-i}$ and consider a
convergent subsequence of $T_i(\bp_{-i}^{(n)})$, with limit $t$.
For every $q_i\in[0,\overline p_i]$, optimality and continuity give
\[
u_i(t,\bp_{-i})
=\lim_n u_i(T_i(\bp_{-i}^{(n)}),\bp_{-i}^{(n)})
\geq\lim_n u_i(q_i,\bp_{-i}^{(n)})
=u_i(q_i,\bp_{-i}),
\]
where the limits are taken along that subsequence.
Uniqueness gives $t=T_i(\bp_{-i})$.
Compactness of the price interval ensures that every subsequence
has a convergent further subsequence, so all limit points agree and
the full sequence converges. Each $T_i$ is continuous.
Brouwer's fixed-point theorem \citep{boothby1971two} applied to
$T=(T_i)_{i=1}^N:\mathcal P\to\mathcal P$ gives a Nash equilibrium.
The centralized maximum also exists by continuity of $R$ on
$\mathcal P$; centralized concavity is not needed here.

Fix a Nash equilibrium $\bx$. Necessary first-order conditions along
feasible coordinate segments imply
\[
\bigl[F_i(\bx)+x_i b_{ii}(\bx)\bigr](p_i-x_i)\leq0
\quad\text{for all }p_i\in[0,\overline p_i].
\]
Writing $\mathbf g=\bm F(\bx)-\bD(\bx)\bx$, summing these
inequalities and choosing $\bp=\mathbf0$ yields
\[
\mathbf g^\top(\bp-\bx)\leq0,
\qquad
\bx^\top\mathbf g\geq0,
\qquad
R(\bx)\geq\bx^\top\bD(\bx)\bx.
\]
These conclusions do not use concavity of the demand functions.

Suppress the argument $\bx$ in $\bB,\bD,\bH$ and $\mathbf C$,
where $\bH=-(\bB+\bB^\top)/2$ and $\mathbf C=\bB+\bD$.
In the following coordinate expressions $b_{ii}=b_{ii}(\bx)$.
The symmetric off-diagonal entries obey
\[
\sum_{j\ne i}|H_{ij}|
\leq\frac12\left(\sum_{j\ne i}|b_{ij}|
                +\sum_{j\ne i}|b_{ji}|\right)
\leq\mu(-b_{ii}).
\]
Hence, for every $\bv$, it holds
\begin{align*}
\bv^\top\bH\bv\geq\sum_i(-b_{ii})v_i^2
   -\frac12\sum_{i\ne j}|H_{ij}|(v_i^2+v_j^2)=\sum_i\left[-b_{ii}-\sum_{j\ne i}|H_{ij}|\right]v_i^2\geq(1-\mu)\bv^\top\bD\bv.
\end{align*}
Define $\bH_\eta=\bH-\eta\bD$. Then
\[
\bH_\eta\succeq(1-\mu-\eta)\bD\succ0,
\]
since $0\leq\eta<1-\mu$.
Diagonalizing $\bD^{-1/2}\bH_\eta\bD^{-1/2}$, whose eigenvalues
are at least $1-\mu-\eta$, shows that
\[
\bH_\eta^{-1}\preceq
\frac1{1-\mu-\eta}\bD^{-1}.
\]
For the interaction term, weighted Cauchy-Schwarz, followed by the
column and row dominance bounds, gives
\begin{align*}
\bv^\top\mathbf C\bD^{-1}\mathbf C^\top\bv
=\sum_i\frac{|\sum_j C_{ji}v_j|^2}{-b_{ii}}\leq\sum_i\frac{\sum_j|C_{ji}|}{-b_{ii}}
                  \sum_j|C_{ji}|v_j^2\leq\mu\sum_jv_j^2\sum_i|C_{ji}|\leq\mu^2\sum_j(-b_{jj})v_j^2.
\end{align*}
Thus, we have that
\[
\mathbf C\bH_\eta^{-1}\mathbf C^\top
\preceq\frac1{1-\mu-\eta}\mathbf C\bD^{-1}\mathbf C^\top
\preceq\frac{\mu^2}{1-\mu-\eta}\bD.
\]
For any $\bp\in\mathcal P$, let $\boldsymbol\Delta=\bp-\bx$ and
$r_i=F_i(\bp)-F_i(\bx)-\nabla F_i(\bx)^\top\boldsymbol\Delta$.
By definition, $E_{\bx}(\bp)=\sum_i p_i r_i$.
Expanding revenue gives the exact identity
\begin{align*}
R(\bp)-R(\bx)
=\boldsymbol\Delta^\top\bm F(\bx)
 +\bx^\top\bB\boldsymbol\Delta
 +\boldsymbol\Delta^\top\bB\boldsymbol\Delta
 +E_{\bx}(\bp)=\mathbf g^\top\boldsymbol\Delta
 +(\mathbf C^\top\bx)^\top\boldsymbol\Delta
 -\boldsymbol\Delta^\top\bH\boldsymbol\Delta
 +E_{\bx}(\bp).
\end{align*}
The equilibrium inequality and the assumed remainder bound imply
\[
R(\bp)-R(\bx)
\leq(\mathbf C^\top\bx)^\top\boldsymbol\Delta
    -\boldsymbol\Delta^\top\bH_\eta\boldsymbol\Delta.
\]
With $\mathbf e=\mathbf C^\top\bx$, the right-hand side equals
\begin{align*}
\frac14\mathbf e^\top\bH_\eta^{-1}\mathbf e\quad-
\left(\boldsymbol\Delta-\frac12\bH_\eta^{-1}\mathbf e\right)^\top
\bH_\eta
\left(\boldsymbol\Delta-\frac12\bH_\eta^{-1}\mathbf e\right)
\leq\frac14\bx^\top\mathbf C\bH_\eta^{-1}\mathbf C^\top\bx.
\end{align*}
At a centralized optimizer, this yields
\begin{align*}
R^*-R(\bx)
\leq\frac14\bx^\top\mathbf C\bH_\eta^{-1}\mathbf C^\top\bx\leq\frac{\mu^2}{4(1-\mu-\eta)}\bx^\top\bD\bx\leq\frac{\mu^2}{4(1-\mu-\eta)}R(\bx).
\end{align*}
Consequently, we have
\[
0<R^*\leq
\left(1+\frac{\mu^2}{4(1-\mu-\eta)}\right)R(\bx).
\]
The multiplier is finite and positive, proving $R(\bx)>0$.
Division now gives
\[
\frac{R(\bx)}{R^*}
\geq\left(1+\frac{\mu^2}{4(1-\mu-\eta)}\right)^{-1}
=\frac{4(1-\mu-\eta)}{4(1-\mu-\eta)+\mu^2}.
\]
If one $\eta$ satisfies the premise at every equilibrium, taking the
infimum over equilibria gives the stated bound for
$\mathrm{PoA}(\bm F,\mathcal P)$.

Each $-b_{ii}(\bp)$ is positive and continuous on the compact box,
so its minimum is positive. Thus the matrix $\underline{\bD}$
defined in the main text is positive definite and satisfies
$\underline{\bD}\preceq\bD(\bx)$ at every $\bx\in\mathcal P$.
Assume $L_i\geq0$,
$\nabla^2F_i(\bp)\preceq L_i\underline{\bD}$ throughout the box,
and $\eta=\frac12\sum_i\overline p_iL_i<1-\mu$.
The segment $\bx+t\boldsymbol\Delta$ lies in $\mathcal P$ for
$0\leq t\leq1$. Taylor's integral formula therefore gives
\begin{align*}
r_i
=\int_0^1(1-t)\boldsymbol\Delta^\top
          \nabla^2F_i(\bx+t\boldsymbol\Delta)
          \boldsymbol\Delta\,dt\leq L_i\left(\int_0^1(1-t)\,dt\right)
           \boldsymbol\Delta^\top\underline{\bD}\boldsymbol\Delta=\frac{L_i}{2}
           \boldsymbol\Delta^\top\underline{\bD}\boldsymbol\Delta.
\end{align*}
Multiplication by $p_i\geq0$ preserves each inequality. Since
$p_i\leq\overline p_i$, summation gives
\begin{align*}
E_{\bx}(\bp)
\leq\frac12\left(\sum_i p_iL_i\right)
          \boldsymbol\Delta^\top\underline{\bD}\boldsymbol\Delta\leq\frac12\left(\sum_i\overline p_iL_i\right)
          \boldsymbol\Delta^\top\underline{\bD}\boldsymbol\Delta\leq\eta\boldsymbol\Delta^\top\bD(\bx)\boldsymbol\Delta.
\end{align*}
This is the required remainder bound uniformly over all feasible
$\bx$ and $\bp$, and in particular at every equilibrium.}

}

\end{APPENDICES}

\end{document}